\pdfoutput=1
\documentclass[a4paper,11pt]{article}
\usepackage{jheppub,bm,tikz,listings}
\usepackage{dcolumn}
\usepackage{subcaption}
\usepackage{hyperref}
\usepackage{float}
\usepackage{amsmath}
\usepackage{amssymb}
\usepackage{amsthm}
\usepackage{mathrsfs}
\usepackage{mathtools}
\usepackage{orcidlink}
\hypersetup{colorlinks=true}
\newcommand{\beq}{\begin{eqnarray}}
\newcommand{\eeq}{\end{eqnarray}}
\newcommand{\non}{\nonumber\\}
\DeclareMathOperator{\U}{U}
\DeclareMathOperator{\SU}{SU}
\DeclareMathOperator{\su}{\mathfrak{su}}

\DeclareMathOperator{\SO}{SO}

\newcommand{\p}{\partial}

\renewcommand{\d}{\mathop{}\!\textup{d}}

\DeclareMathOperator{\diag}{\rm diag}

\DeclareMathOperator{\Tr}{Tr}
\DeclareMathOperator{\Id}{Id}

\newcommand{\CP}{\mathbb{C}P}

\newcommand{\bphi}{\bm{\phi}}

\newcommand{\bea}{\begin{eqnarray}}
\newcommand{\eea}{\end{eqnarray}}
\newcommand{\be}{\begin{equation}}
\newcommand{\ee}{\end{equation}}

\usetikzlibrary{shapes.geometric, arrows}
\tikzstyle{module} = [rectangle, rounded corners, 
minimum width=3cm,
minimum height=1cm,
text centered,
draw=black,
fill=red!30]
\tikzstyle{header} = [rectangle, 
minimum width=3cm,
minimum height=1cm,
text centered,
draw=black,
fill=orange!30]
\tikzstyle{arrow} = [thick,->,>=stealth]

\title{\texttt{cuSkyrmion}: A CUDA--OpenGL framework for interactive simulation and visualization of nuclei as Skyrmions}
\author[a,b]{Sven Bjarke Gudnason\orcidlink{0000-0001-9255-5940},}
\author[c]{Paul Leask\orcidlink{0000-0002-6012-0034}}
\affiliation[a]{Institute of Contemporary Mathematics, School of
  Mathematics and Statistics, Henan University, Kaifeng, Henan 475004,
  P.~R.~China}
\affiliation[b]{Department of Physics, Chemistry and Pharmacy,
  University of Southern Denmark, Campusvej 55, 5230 Odense M,
  Denmark}
\affiliation[c]{Department of Physics, KTH Royal Institute of
  Technology, 10691 Stockholm, Sweden}
\emailAdd{gudnason@henu.edu.cn}
\emailAdd{palea@kth.se}

\abstract{
We introduce \texttt{cuSkyrmion}, a 3-dimensional Skyrme model computation and visualization software, that is written in \texttt{CUDA C} for rapid computation and visualization of especially the arrested Newton flow algorithm.
The programme is interactive and lets the user construct Skyrmions either with configuration files, specifying coordinates, or simply in run-time using the keyboard and mouse.
Rational map ansatz constituent Skyrmions can be inserted at any time and a random generator can produce a stochastic initial configuration.
The software is composed into three main modules being a computational module, a rendering module and a main programme.
The rendering/visualization module can readily be used by other computational modules and a \texttt{Python}-fork, \texttt{skyrmion\_solver}, has been developed demonstrating the re-usability of the code.

\newpage
\bigskip\noindent
{\sc Program Summary}\\
Program title: \texttt{cuSkyrmion}\\
CPC Library link to program files: (to be added by Technical Editor)\\
Developer's respository link: \href{https://bitbucket.org/sbgudnason/cuskyrmion}{https://bitbucket.org/sbgudnason/cuskyrmion}\\
Licensing provisions: MIT license (MIT)\\
Programming language: CUDA C/C++\\
Nature of problem (approx. 50-250 words):
Three-dimensional Skyrmions are topological solitons in non-linear
differential classical field equations and require costly numerical
calculations to find, especially with high accuracy of energy and
inertia tensors.
The non-linear field equations' solutions are solitons whose fields
on a three-dimensional simulation box are mapped to the three-sphere.
The number of solutions grows perhaps exponentially with the
topological degree [S1, Fig.~11a] and the need for a visualizing and fast
computation package is crucial for further research in this research
field. \\
Solution method (approx. 50-250 words):
\texttt{cuSkyrmion} solves the non-linear numerical field equations
using the so-called arrested-Newton flow and a fourth-order
finite-difference approximation for the derivatives of the partial
differential equations.
The unique improvement over earlier packages [S2], is that
\texttt{cuSkyrmion} is fully implemented in CUDA, runs entirely on
the GPU and comes with a built-in visualizer that performs ray-tracing
of the field configuration at the current step of the calculation.
This enables the researcher to see what the fields are doing, what is
attracting and what is collapsing and what outcome may form, before
waiting for a long calculation to finish.
\texttt{cuSkyrmion} comes with the unique ability to modify the
fields during run-time by inserting more solitons using the rational
map approximation and the product ansatz on-the-go.\\
Additional comments including Restrictions and Unusual features (approx. 50-250 words):
The code is split into a computational module, a visualization module and a main programme. But the computational module is designed for $\SU(2)$ fields using a 4-vector formulation and is not readily changeable to different field theories with different field content.
However, we describe and link to the Python-fork of the programme [S3] that has a modular structure in the computational module, so that a new theory can easily be implemented with limited programming experience, since the solver module is independent of the theory module. \\
References:\\
\phantom{.}[S1] S.~B.~Gudnason, C.~Halcrow,
``A Sm\o rg\aa sbord of Skyrmions'',
\href{http://dx.doi.org/10.1007/JHEP08(2022)117}{JHEP 2208, 117 (2022)} [\href{http://arxiv.org/abs/2202.01792}{arXiv:2202.01792 [hep-th]}].\\
\phantom{.}[S2] \href{https://github.com/chrishalcrow/Skyrmions3D.jl}{https://github.com/chrishalcrow/Skyrmions3D.jl}\\
\phantom{.}[S3] \href{https://github.com/Paulnleask/skyrmion_solver}{https://github.com/Paulnleask/skyrmion\_solver}
}

\begin{document}
\maketitle


\section{Introduction}\label{sec:intro}

Topological solitons play a central role in a wide range of nonlinear field theories, providing stable, particle-like excitations whose existence is guaranteed by topology rather than by linear stability.
In nuclear and hadronic physics, the Skyrme model occupies a distinguished position as an effective low-energy description of baryons and nuclei, in which the baryon number is identified with a topological charge \cite{Manton_Sutcliffe_2004,Manton_2022}.
Within this framework, classical field configurations known as Skyrmions encode many qualitative and quantitative features of nuclear structure, including binding energies, shapes, and rotational spectra.

Despite its conceptual simplicity, the Skyrme model gives rise to a highly non-linear energy functional whose minimization presents significant analytical and numerical challenges, particularly for higher baryon numbers.
While powerful approximation schemes such as the rational map ansatz capture much of the qualitative structure of low-charge Skyrmions \cite{Houghton_1998}, fully relaxed numerical solutions are essential for accurate determination of energies, inertia tensors, and other physical observables.
This is especially true for multi-Skyrmion configurations, where the energy landscape becomes increasingly complex and populated by many local minima corresponding to different clusterings and symmetry types \cite{Gudnason_2022}.

In recent years, the growing interest in mechanical and energy--momentum tensor properties of solitons has further motivated high-precision numerical studies.
Observables such as stress distributions, quadrupole moments, inertia tensors, and the monopole D-term provide detailed information about the internal structure of Skyrmions and enable direct comparison with modern descriptions of hadrons and nuclei.
Reliable computation of these quantities requires not only accurate solutions of the field equations, but also careful treatment of translational invariance, centre-of-mass effects, and numerical convergence diagnostics such as virial constraints.

In parallel with these theoretical developments, advances in heterogeneous computing have made graphics processing units (GPUs) an increasingly powerful platform for large-scale field-theoretic simulations.
However, existing numerical implementations of the Skyrme model are typically CPU-based and optimized primarily for offline relaxation, limiting both interactivity and the ability to explore solution spaces dynamically.
In particular, to the best of our knowledge, no existing software
provides real-time visualization of Skyrme-field evolution and
associated observables during energy minimization.

In this work we present \texttt{cuSkyrmion}, a CUDA-accelerated numerical framework for computing static Skyrmion solutions and their associated physical observables.\footnote{Refs.~\cite{Bacciu:2025wbr,DellaTorre:2026stp} also solve a partial
  differential equation (the Parker transport equation) using
  a CUDA-accelerated numerical code. The problem addressed in those
  papers require too large a simulation domain (particle dynamics on
  cosmic scales) and so uses a stochastic method essentially reducing
  the PDE problem to one ordinary differential equation (ODE) on one
  CUDA thread and independent from the other threads, which differs
  significantly from our approach to solving the PDE on a box (where
  the threads do not decouple).
  Finally, the code of Refs.~\cite{Bacciu:2025wbr,DellaTorre:2026stp}
  does not provide a continuous visualization via ray-tracing through a
  field configuration during the calculation, like \texttt{cuSkyrmion}
  does.
}
The code implements a high-order finite-difference discretization of the Skyrme model in the sigma-model (vector) formulation, together with an efficient arrested Newton flow minimization scheme.
This second-order relaxation method accelerates convergence in stiff energy landscapes while maintaining stability through an energy-based arrest criterion, making it particularly well suited to multi-Skyrmion relaxation.
The entire minimization procedure is executed on the GPU, allowing for efficient scaling with lattice size and rapid convergence even for higher baryon numbers.

A distinguishing feature of \texttt{cuSkyrmion} is the integration of real-time visualization with the numerical solver.
During relaxation, quantities such as baryon density, energy density, and selected field components are rendered interactively using GPU-based rendering pipelines.
This enables direct, real-time inspection of Skyrmion formation, deformation, and symmetry breaking as the flow proceeds, providing both qualitative insight and immediate diagnostic feedback.
Such real-time visualization is particularly valuable for identifying metastable configurations, diagnosing convergence issues, and exploring families of initial conditions.

Another distinguishing feature of \texttt{cuSkyrmion} is the ability to modify the Skyrmion configuration interactively, in addition to simply starting with an initial configuration as a starting point.
The sm\"org\aa sbord's random generation algorithm is also included, allowing one to add randomly placed and randomly oriented 1-Skyrmions on top of the current configuration, be it empty or containing already Skyrmions.
A particularly useful feature is the possibility to add rational map Skyrmions with charge $B$ equal to 1 through 9 to the current Skyrmion configuration at a desired position with a desired orientation.
The process of adding the rational map Skyrmion lets the user move the new addition (say a 2-Skyrmion) around in the configuration space with the visualizer continuously showing what the product ansatz does to the current configuration multiplied by the new Skyrmion.
Once the position, the orientation and the isospin orientation has been chosen by the user, the energy minimizing arrested Newton flow algorithm relaxes the configuration to the nearest local minimum.
This takes the creation of multi-Skyrmions closer to playing a game than figuring out multiplication and rotation matrices in a construction of an initial configuration.
With the immediate visualization, the user can quickly see if the result is converging towards a desired state or one can quickly start over and try again or try something else.
\texttt{cuSkyrmion} can also be loaded with an existing datafile, that is either a previously saved state from \texttt{cuSkyrmion} (in tab-separated text  or binary format) or a tab-separated text  file created using third-party package (e.g.~python, Mathematica, MATLAB, or other), or a configuration file can be loaded that places $N$ $B$-Skyrmions at specified positions with specified orientations and isospin orientation.

Beyond the computation of energies and baryon densities, we focus on the extraction of physically meaningful observables from the relaxed solutions.
These include the centre of mass, root-mean-square radius, electric quadrupole tensor, inertia tensors associated with spatial and isospin rotations, and stress-tensor derived quantities such as the virial constraint and the D-term.
Special care is taken to evaluate coordinate-dependent observables in the centre-of-mass frame, ensuring that translational artefacts are removed.
Together, these quantities provide a detailed characterization of Skyrmion structure and furnish inputs for semi-classical quantization and related analyses.
The centre-of-mass position can also be used to check that the final Skyrmion configuration is not accidentally close to the boundary of the simulation area.

The paper is organized as follows.
In Sec.~\ref{sec: Skyrme model} we review the Skyrme model and its formulation in terms of a constrained vector field.
Sec.~\ref{sec: Arrested Newton flow} describes the numerical discretization and the arrested Newton flow minimization algorithm.
Sec.~\ref{sec: Constructing multiskyrmion configurations} describes the methods for creating multi-Skyrmion configurations.
In Sec.~\ref{sec: Physical properties of skyrmions} we define the physical observables considered in this work and discuss their numerical evaluation.
In Sec.~\ref{sec: Program design and implementation in CUDA C} the programme design is briefly explained and in Sec.~\ref{sec: User manual} we provide a user manual for the software. Neat examples are then illustrated in Sec.~\ref{sec: Neat examples}.
The programme is then benchmarked and scaling in function of the number of CUDA cores is analysed in Sec.~\ref{sec:benchmark}.
The Python-fork \texttt{skyrmion\_solver} is presented in
Sec.~\ref{sec: The Python port: skyrmion_solver}.
Finally, we conclude with an outlook in Sec.~\ref{sec: Outlook}.


\section{The Skyrme model}
\label{sec: Skyrme model}

The Skyrme model was introduced by Tony Skyrme in the early 1960'ies
as a simple model for the nucleon \cite{Skyrme_1961}. 
The massive Skyrme model consists of a single scalar field $U(t,\mathbf{x}) \in \SU(2)$, and is defined by the Lagrangian density
\begin{equation}
\label{eq: Skyrme Lagrangian}
    \mathcal{L} = \frac{F_\pi^2}{16\hbar} \Tr(L_\mu L^\mu) + \frac{\hbar}{32e^2} \Tr\left( \left[L_\mu, L_\nu\right] \left[L^\mu, L^\nu \right] \right) - \frac{1}{8\hbar^3}F_\pi^2 m_\pi^2 \Tr(\Id_2 - U),
\end{equation}
where $F_\pi$ is the pion decay constant, $m_\pi$ is the pion mass, $e$ is a dimensionless parameter called the Skyrme coupling, and $\hbar$ is the reduced Planck constant.
Let us denote the Lie algebra of $\SU(2)$ by $\su(2)$.
Then, the pull-back of the left Maurer-Cartan form $\omega$ defines the $\su(2)$-valued left current $L_\mu = U^\dagger \partial_\mu U$, where $\mu=0,1,2,3$ is the spacetime index that is raised with the mostly-positive Minkowski metric.

Using energy and length units of
\beq
\tilde{E}=F_\pi/4e \textrm{(MeV)},\qquad
\tilde{L}=2\hbar/e F_\pi \textrm{(fm)},
\label{eq:energy_length_units}
\eeq
the Lagrangian can be expressed in the dimensionless form
\begin{align}
    L = \int_{\mathbb{R}^3} \textup{d}^3x \left\{ \frac{1}{2} \Tr(L_\mu L^\mu) + \frac{1}{16} \Tr\left( \left[L_\mu, L_\nu\right] \left[L^\mu, L^\nu\right] \right) -m^2 \Tr\left( \Id_2 - U \right) \right\},
    \label{eq:Lag_dimensionless}
\end{align}
where $m = 2 m_\pi/F_\pi e$ is the rescaled pion mass.
The pion mass potential explicitly breaks the chiral $\SO(4)$ symmetry of the model to an  $\SO(3)\cong\SU(2)$ isospin symmetry, given by the conjugation $U \mapsto AU A^\dagger$ with $A \in \SU(2)$.
Upon quantization, this gives rise to the quantity that distinguishes protons and neutrons: isospin.

The static energy functional is obtained from the potential part of the Lagrangian:
\begin{align}
    E = \, & \int_{\mathbb{R}^3} \textup{d}^3x \left\{ m^2 \Tr\left( \Id_2 - U \right) -\frac{1}{2} \Tr(L_i L_i) - \frac{1}{16} \Tr\left( [L_i, L_j] [L_i, L_j] \right) \right\},
\label{eq: Skyrme model - Static energy}
\end{align}
where the lower-case Latin indices $i,j,k=1,2,3$ correspond to only spatial dimensions. 
A field configuration $U$ which minimizes the static energy functional \eqref{eq: Skyrme model - Static energy} is referred to as a \emph{Skyrmion} and the static energy $E$ is often interpreted as the classical mass of the Skyrmion.

For field configurations to have finite energy, they must satisfy the boundary condition $U\rightarrow\Id_2$ as $|\mathbf{x}| \rightarrow \infty$.
This yields a one-point compactification of the domain $\mathbb{R}^3 \cup \{\infty\} \cong S^3$, such that topologically $U: S^3 \rightarrow S^3$ at a fixed time.
The disjoint homotopy classes of such maps are labelled by their
topological degree $B \in \pi_3(S^3)=\mathbb{Z}$ and the fields are
necessarily topologically stable configurations\footnote{The
topological stability assumes the boundary condition
$U\rightarrow\Id_2$. On a finite simulation domain (e.g.~a box in
$\mathbb{R}^3$) topological stability requires the imposition of
Dirichlet boundary conditions on the boundary of the domain. }.
The topological degree is identified with the physical baryon number.
So, we often to refer to $B$ as the baryon number, which may be computed using
\begin{equation}
\label{eq: Baryon number}
    B = \int_{\mathbb{R}^3} \textup{d}^3x \, \mathcal{B}^0,
\end{equation}
where
\begin{equation}
\label{eq: Baryon density}
    \mathcal{B}^\mu = \frac{1}{24\pi^2} \epsilon^{\mu \nu \rho \sigma} \Tr(L_\nu L_\rho L_\sigma),
\end{equation}
is the baryon current density.

The Euler--Lagrange equation is obtained by varying $U$ through $U \mapsto U e^\varepsilon$ with $\varepsilon : \mathbb{R}^3 \to \su(2)$.
Writing $\delta U = U\varepsilon$, one finds the static field equation
\begin{equation}
\label{eq:skyrme-static-eom}
    \partial_i
    \left(
        L_i + \frac{1}{4}[L_j,[L_i,L_j]]
    \right)
    - \frac{m^2}{2}(U-U^\dagger) = 0,
\end{equation}
where the mass term contributes only with the traceless $\su(2)$ part of $U$.


\subsection{Sigma-model formulation}

For numerical work it is convenient\footnote{This can be seen from the fact that computing the equation of motion of the field $U$ yields the equation of motion for $\phi$ twice: mathematically this is no problem, but numerically this is redundant. } to represent the Skyrme field $U:\mathbb{R}^3\to \SU(2)$ by a unit four-vector $\bphi=(\phi^0,\phi^1,\phi^2,\phi^3):\mathbb{R}^3\to S^3\subset\mathbb{R}^4$,
via
\begin{equation}
    U(\mathbf{x}) = \phi_0(\mathbf{x})\,\Id_2 + i \phi^a(\mathbf{x})\tau^a, \quad \bphi\cdot\bphi = 1.
\end{equation}
The point-wise constraint $\bphi\cdot\bphi=1$ ensures $U(\mathbf{x})\in \SU(2)$ everywhere and $\tau^a$ are the Pauli spin matrices.

In dimensionless Skyrme units, the static energy density may be written in the more convenient non-linear sigma-model form,
\begin{equation}
\label{eq:SkyrmeEnergy_phi}
    \mathcal{E} = \partial_i\bphi\cdot\partial_i\bphi + \frac{1}{2}\left[ (\partial_i\bphi\cdot\partial_i\bphi) (\partial_j\bphi\cdot\partial_j\bphi) - (\partial_i\bphi\cdot\partial_j\bphi) (\partial_i\bphi\cdot\partial_j\bphi) \right] + 2m^2\,(1-\phi^0).
\end{equation}
This is equivalent to the usual $\SU(2)$ form of the Skyrme energy, but avoids explicit matrix operations in the discretization.
Similarly, the baryon current density can be expressed as
\begin{equation}
    \mathcal{B}^0 = \frac{1}{12\pi^2}\, \epsilon^{ijk}\epsilon_{ABCD}\, \phi^A\, \partial_i\phi^B\, \partial_j\phi^C\, \partial_k\phi^D, \quad A=0,1,2,3.
\label{eq:B0_components}
\end{equation}


\section{Arrested Newton flow}
\label{sec: Arrested Newton flow}

To obtain static solutions of the field equations, we minimize the energy functional numerically using the arrested Newton flow method, which is a second-order relaxation scheme in a fictitious time variable.
The field is written as a four-component real vector $\bphi=\{\phi^{0},\phi^{1},\phi^{2},\phi^{3}\}$ satisfying the point-wise constraint $\bphi\cdot\bphi=1$, so that $\bphi$ takes values in $S^{3}$.
The method proceeds by evolving $\bphi(\mathbf{x},t)$ according to a second-order-in-time flow equation derived from the functional derivative of the static energy,
\begin{equation}
    \frac{\partial^2\phi^A}{\partial t^2} = -\frac{\delta \mathcal{E}}{\delta \phi^A}
    +\frac{\delta \mathcal{E}}{\delta \phi^B}\phi^B\phi^A,
    \label{eq:Newton_flow}
\end{equation}
where the Euler-Lagrange field equation
\begin{align}
    \frac{\delta \mathcal{E}}{\delta \phi^A} &=  -2 \Big\{ \partial_{ii}\phi^A \left[1 + \left( \partial_j \phi^B \right)^2 \right] + \partial_i\phi^A \left( \partial_{ij}\phi^B\,\partial_j\phi^B - \partial_{jj}\phi^B\,\partial_i\phi^B \right) - \partial_{ij}\phi^A \left( \partial_i\phi^B\,\partial_j\phi^B \right) \nonumber \\
    &\phantom{=-2\Big\{\ }
    + m^2\delta^A_0\Big\},
\label{eq: Sigma model EL equations}
\end{align}
and $\partial_{ij}$ is a short-hand for the double partial derivative $\partial_{ij}:=\partial_i\partial_j$.
The last term in Eq.~\eqref{eq:Newton_flow} ensures mathematically that the field $\bphi$ retains its unit length constraint (assuming it has a unit length in the initial condition)\footnote{Without this term, even the vacuum does not correspond to a fixed point of the equations of motion: it is thus crucial for finding solutions in the non-linear sigma model. }.
We will, however, also impose the unit length constraint at every step to ensure no numerical inaccuracies violate the non-linear sigma model unit length constraint.

The evolution of Eq.~\eqref{eq:Newton_flow} is interpreted as a fictitious dynamical system whose stationary points correspond to critical points of the energy functional.
In contrast to first-order gradient flow, the inclusion of second-order time derivatives significantly accelerates relaxation toward local minima while still allowing the system to escape shallow directions in configuration space.

During the evolution, the potential energy associated with the fictitious time dynamics is monitored.
Whenever the potential energy begins to grow, indicating that the flow is overshooting a minimum or becoming unstable, the evolution is arrested by setting the time derivatives $\partial_{t}\bphi$ to zero while keeping the field configuration itself fixed.
The evolution is then restarted from this configuration with zero velocity.
This arresting procedure is applied repeatedly throughout the relaxation and ensures numerical stability while retaining the rapid convergence properties of second-order flow.
The process is continued until 
the magnitude of the functional derivative $\delta E/\delta\phi$ falls below prescribed tolerances, at which point the configuration is taken to approximate a static solution.

The constraint $|\bphi|=1$ is enforced numerically by explicit projection after each update of the field, replacing $\bphi$ by $\bphi/|\bphi|$ at every lattice site.
This simple procedure is sufficient to maintain the constraint to high accuracy throughout the evolution.

The spatial domain is discretized on a uniform cubic lattice approximating $\mathbb{R}^{3}$, with either periodic boundary conditions or sufficiently large boxes together with vacuum boundary conditions, depending on the physical problem under consideration.
Spatial derivatives appearing in the energy functional and its variation are approximated using fourth-order finite difference schemes.
For a lattice spacing $h$, the first derivative of a field component $\phi^A$ in the $x$-direction is approximated by
\begin{equation}
    \partial_{x}\bphi_{i,j,k} \approx \frac{1}{h_x}\left( -\frac{1}{12}\bphi_{i+2,j,k} + \frac{2}{3}\bphi_{i+1,j,k} - \frac{2}{3}\bphi_{i-1,j,k} + \frac{1}{12}\bphi_{i-2,j,k} \right),
\end{equation}
with $i,j,k$ specifying the lattice site, $h_x$ being the lattice spacing in the $x$-direction
and the second derivative is approximated by
\begin{equation}
    \partial_{x}^{2}\bphi_{i,j,k} \approx \frac{1}{h_x^2} \left( -\frac{1}{12}\bphi_{i+2,j,k} + \frac{4}{3}\bphi_{i+1,j,k} - \frac{5}{2}\bphi_{i,j,k} + \frac{4}{3}\bphi_{i-1,j,k} - \frac{1}{12}\bphi_{i-2,j,k} \right),
\end{equation}
with analogous expressions used in the $y$ and $z$ directions.
All terms in the energy functional are constructed consistently using these fourth-order approximations.
We can then regard the static energy as a function $E:\mathcal{C} \rightarrow \mathbb{R}$, where the discretized configuration space is the manifold $\mathcal{C}=(S^3)^{N^3} \subset \mathbb{R}^{4N^3}$.
This choice substantially reduces discretization errors compared to second-order schemes and is essential for resolving the detailed structure of three-dimensional Skyrme field configurations.

The arrested Newton flow method combines the simplicity of explicit relaxation schemes with improved convergence properties and has proven effective for computing static solitons in Skyrme-type models in both two and three spatial dimensions.
It requires no linear solves or matrix inversions, is straightforward to implement, and is well suited to large-scale computations on uniform grids.


\section{Constructing multi-Skyrmion configurations}
\label{sec: Constructing multiskyrmion configurations}

The \texttt{cuSkyrmion} package comes with built-in Skyrmion components that may be inserted at run-time in the simulation or via a configuration file, which enables the user to construct arbitrary multi-Skyrmions at will.


\subsection{Rational map ansatz}
\label{subsec: Rational map ansatz}

The rational map ansatz provides an efficient approximation to Skyrmion solutions by separating radial and angular dependence.
It exploits the fact that finite-energy Skyrme fields satisfy the boundary condition $U(\mathbf{x}) \rightarrow \Id_2$ as $|\mathbf{x}|\rightarrow\infty$,
so that static configurations may be viewed as maps $U : S^3 \rightarrow S^3$,
classified by their degree $B$, the baryon number, and then it assumes that the radial suspension is a good approximation to the true Skyrmion solution such that at every radius $r$ from the origin the Skyrmion is described by the same degree $B$ rational map $S^2\to S^2$.

Introducing spherical coordinates $(r,\theta,\phi)$, the angular dependence of the field may be described using the Riemann sphere coordinate
\begin{equation}
    z = \tan\left(\frac{\theta}{2}\right)e^{i\phi}.
\end{equation}
A rational map is a holomorphic map between Riemann spheres, $R : S^2 \rightarrow S^2$, given by
\begin{equation}
    R(z) = \frac{p(z)}{q(z)},
\end{equation}
where $p(z)$ and $q(z)$ are complex polynomials with no common factors. The degree of the rational map is
\begin{equation}
    \deg R = \max\{\deg p,\deg q\} = B,
\end{equation}
which coincides with the baryon number of the associated Skyrmion configuration.
The rational map ansatz then takes the form
\begin{equation}
    U(r,z) = \exp\!\left( i f(r)\, \hat{\mathbf{n}}_R(z)\cdot \boldsymbol{\tau} \right),
\end{equation}
where $f(r)$ is a radial profile function satisfying the boundary conditions $f(0)=\pi$, $f(\infty)=0$ and
\begin{equation}
    \hat{\mathbf{n}}_R(z) = \frac{1}{1+|R|^2} \left( 2\,\textup{Re}\,R,\, 2\,\textup{Im}\,R,\, 1-|R|^2 \right),
\end{equation}
is the unit vector on $S^2$ determined by the rational map.
The angular structure of the Skyrmion is therefore entirely encoded in $R(z)$, while the radial dependence is controlled by $f(r)$.

An important feature of the rational map ansatz is that the symmetries of the Skyrmion are inherited from the symmetries of the rational map.
If a spatial rotation acts on the domain sphere and can be compensated by a Möbius transformation of the target sphere leaving $R(z)$ invariant, then the corresponding Skyrmion possesses that symmetry.
For baryon numbers $B=1,\dots,8$, the minimal-energy rational maps exhibit the symmetry groups detailed in Tab.~\ref{tab: RMA}. 
Here $T_d$, $O_h$, and $Y_h$ denote the tetrahedral, octahedral (cubic), and icosahedral symmetry groups respectively, while $D_{nd}$ and $D_{\infty h}$ denote dihedral symmetry groups (for a good reference for discrete symmetry groups, see Ref.~\cite{Jacobs:2005}).
These symmetries closely match those observed in fully relaxed numerical Skyrmion solutions and explain the emergence of polyhedral (fullerene-like) structures at higher baryon number (when the pion mass vanishes).

\begin{table}
\centering
    \begin{tabular}{c c c c}
        \hline
        $B$ & Symmetry Group & $p(z)$ & $q(z)$ \\
        \hline
        1 & $O(3)$ & $z$ & $1$ \\
        2 & $D_{\infty h}$ & $z^2$ & $1$ \\
        3 & $T_d$ & $\sqrt{3}az^2-1$ & $z(z^2-\sqrt{3}a)$ \\
        4 & $O_h$ & $z^4 + 2\sqrt{3}iz^2 + 1$ & $z^4 -2\sqrt{3}iz^2 + 1$ \\
        5 & $D_{2d}$ & $z(z^4 + bz^2 + a)$ & $az^4 - bz^2 + 1$ \\
        6 & $D_{4d}$ & $z^4-a$ & $z^2(az^4+1)$ \\
        7 & $Y_h$ & $bz^6-7z^4-bz^2-1$ & $z(z^6+bz^4+7z^2-b)$ \\
        8 & $D_{6d}$ & $z^6-a$ & $z^2(az^6+1)$ \\
        \hline
    \end{tabular}
    \caption{Rational maps $R(z)=p(z)/q(z)$, and their associated symmetry groups, for Skyrmions up to baryon number $B=8$.
    For the values of the constants $a$ and $b$, see Ref.~\cite{Houghton_1998}.
    }
    \label{tab: RMA}
\end{table}


\subsection{The sm\"org\aa sbord ansatz}
\label{subsec: The smorgasbord ansatz}

While the rational map ansatz provides accurate approximations to single, highly symmetric Skyrmions, configurations describing separated or clustered solitons are more naturally constructed using the product ansatz. 
Let $U_1(\mathbf{x})$ and $U_2(\mathbf{x})$ be two Skyrme fields with baryon numbers $B_1$ and $B_2$, respectively.
The (non-symmetrized) product ansatz constructs a configuration of total baryon number $B=B_1+B_2$ by superposing the fields multiplicatively,
\begin{equation}
    U(\mathbf{x}) = U_1(\mathbf{x}-\mathbf{X}_1)\, A\, U_2(\mathbf{x}-\mathbf{X}_2)\, A^\dagger,
\end{equation}
where $\mathbf{X}_1$ and $\mathbf{X}_2$ are spatial translation vectors and $A \in \SU(2)$ represents a relative isorotation.
More generally, for $N$ constituent Skyrmions,
\begin{equation}
\label{eq: Product ansatz}
    U(\mathbf{x}) = \prod_{k=1}^{N} A_k\, U_k(\mathbf{x}-\mathbf{X}_k)\, A_k^\dagger,
\end{equation}
which includes also an overall isorotation.
Because the Skyrme field takes values in $\SU(2)$, multiplication preserves the boundary condition $U \to \Id_2$ at spatial infinity, and the total baryon number is additive.

The product ansatz is exact only when solitons are infinitely separated, but it provides an excellent initial condition for numerical relaxation.
It allows one to construct multi-soliton configurations with prescribed spatial arrangement and relative isospin orientations.
These relative orientations are crucial, as the interaction energy between Skyrmions depends sensitively on their isorotational alignment.

A systematic implementation of this idea is provided by the sm\"org\aa sbord method \cite{Gudnason_2022}.
The sm\"org\aa sbord approach generates a large family of initial conditions by combining 
$B$ 1-Skyrmions 
using the product ansatz with varying relative positions and isospin orientations. The method proceeds as follows:

\begin{enumerate}
    \item The constituent 1-Skyrmions are placed at prescribed spatial locations $\mathbf{X}_k$, generated randomly, but with a constraint on the maximal distance from one constituent to at least one of the others.
    \item Independent isorotations $A_k \in \SU(2)$ are assigned to each constituent, sampling distinct relative orientations.
    \item The fields are combined multiplicatively using the product ansatz \eqref{eq: Product ansatz} to form a composite configuration.
    \item The resulting field is numerically relaxed using the arrested Newton flow method
    to obtain a nearby local minimum of the energy functional.
\end{enumerate}

By systematically 
varying spatial arrangements, and relative isorotations, the sm\"org\aa s\-bord method explores a wide landscape of candidate configurations for a given baryon number.
This approach is particularly effective for intermediate and higher charges, where the energy landscape contains many metastable cluster configurations and where direct construction via rational maps may not capture all relevant local minima.

In practice, the product ansatz serves as the mechanism that encodes both spatial clustering and isospin alignment in the initial data.
The subsequent numerical relaxation accounts for non-linear interactions between constituents and yields physically meaningful multi-Skyrmion solutions.
The method therefore provides a complementary construction to the rational map ansatz, favouring cluster-based structures over highly symmetric single-shell configurations.
This is of particular importance when the pion mass potential is included, as the fullerene-like structures are unstable and collapse to coalesced multi-Skyrmion configurations with less symmetry.

A comment is in store about the product ansatz.
The order of the Skyrmions matters in the product ansatz as it is not commutative.
Indeed, a symmetric product ansatz has been suggested in the literature and may describe certain clusters closer to their local minimum.
However, the advantage of the asymmetric product ansatz is that the sigma model constraint is automatically satisfied and that the baryon number is exactly the sum of constituent baryon numbers.
Indeed, the problem of non-commutativity is made up for by randomly producing many initial states in the sm\"org\aa sbord programme.


\section{Physical properties of Skyrmions}
\label{sec: Physical properties of skyrmions}

A key ingredient in the definition of physical observables in the Skyrme model is the energy--momentum tensor.
In dimensionless Skyrme units, the energy--momentum tensor is defined by
\begin{align}
\label{eq: Stress tensor}
    T_{\mu\nu} = -\Tr(L_\mu L_\nu) - \frac{1}{4} \eta^{\alpha\beta} \Tr\left( [L_\mu,L_\alpha][L_\nu,L_\beta] \right) + \eta_{\mu\nu}\mathcal{L},
\end{align}
where $L_\mu = U^\dagger \partial_\mu U \in \mathfrak{su}(2)$ is the left-invariant current, $\eta_{\mu\nu}=\textup{diag}(-1,1,1,1)$ is the Minkowski metric with the mostly-positive signature, and $\mathcal{L}$ is the Skyrme Lagrangian density.
This tensor follows from Noether’s theorem applied to spacetime translations and encodes the local energy density, momentum density, and internal stresses of the field configuration.
In particular, $T_{00}$ represents the energy density, $T_{0i}$ the momentum density, and $T_{ij}$ the spatial stress tensor, which characterizes the internal force distribution within the soliton.

For static field configurations, $\partial_0 U = 0$, so that $L_0 = 0$ and the momentum density vanishes.
In this case the timelike component reduces to the static energy density, $T_{00}=\mathcal{E}_{\textup{stat}}$, and the total energy is obtained by spatial integration.


\subsection{Centre of mass}

From the baryon density \eqref{eq: Baryon density}, we define the centre of mass of a Skyrmion by
\begin{equation}
\label{eq: CoM}
    \mathbf{X}_c = \frac{1}{B} \int_{\mathbb{R}^3}\textup{d}^3x\, \mathbf{x}\,\mathcal{B}^0(\mathbf{x}),
\end{equation}
where $B$ is the baryon number.
This defines the centre of the physical system and provides the natural origin with respect to which spatial moments of physical observables are defined.

It is assumed that multi-Skyrmions are connected and no separated clusters exist. In case one has ended up with spatially separated clusters, one needs to restart the computations from different initial conditions or add in extra Skyrmions to connect the clusters -- that is, if one desires to compute quantitative observables.

Quantities that explicitly depend on spatial coordinates, such as inertia tensors, multipole moments, and moments of the energy--momentum tensor (including the D-term), must be evaluated relative to this centre in order to eliminate spurious contributions arising from translational 
of the soliton.
We therefore introduce centre-of-mass coordinates
\begin{equation}
    \mathbf{r} = \mathbf{x} - \mathbf{X}_c,
\end{equation}
which are used throughout when evaluating coordinate-dependent observables.
By contrast, local densities such as the baryon density and the stress tensor itself are translationally invariant and are unaffected by shifts of the coordinate origin.


\subsection{Skyrmion size}

A natural measure of the spatial extent of a Skyrmion is provided by its root-mean-square (RMS) radius, defined by \cite{Adam_2016}
\begin{equation}
\label{eq: RMS radius}
    R_{\textup{Sk}} = \left( \frac{1}{B} \int_{\mathbb{R}^3}\textup{d}^3x\, |\mathbf{r}|^2\,\mathcal{B}^0(\mathbf{r}) \right)^{\frac{1}{2}}.
\end{equation}
This quantity characterizes the typical spatial size of the baryon density distribution and is evaluated relative to the centre of mass.


\subsection{Moments of inertia}

The rotational properties of a Skyrmion are encoded in a set of inertia tensors that characterize its response to rotations in physical space and isospace.
These tensors contain intrinsic properties of a given static field configuration and are determined entirely by the spatial distribution of the Skyrme field and its currents.
A static Skyrmion configuration $U(\mathbf{x})$ is not unique, but belongs to a family of energetically degenerate configurations related by spatial translations, spatial rotations, and isorotations.
Working in the centre-of-mass frame $\mathbf{r}=\mathbf{x}-\mathbf{X}_c$, we neglect translational degrees of freedom and focus exclusively on rotational properties.
Spatial rotations are represented by elements of $\SO(3)_J$, while isorotations act in the internal $\SO(3)_I$ symmetry space of the model.
Both actions are conveniently represented using $\SU(2)$ matrices.
The homomorphism $D:\SU(2)\to\SO(3)$ is given by \cite{Manko_2007}
\begin{equation}
    D(B)_{ij} = \frac{1}{2} \Tr\!\big( \tau^i B \tau^j B^\dagger \big),
\end{equation}
so that a spatial rotation acts on coordinates as $\mathbf{r}\mapsto D(B)\mathbf{r}$ and induces the pull-back
\begin{equation}
    U(\mathbf{r}) \mapsto U\!\left(D(B)^{-1}\mathbf{r}\right).
\end{equation}
Isorotations act directly on the field according to
\begin{equation}
    U(\mathbf{r}) \mapsto A\,U(\mathbf{r})\,A^\dagger,
\end{equation}
with $A\in\SU(2)$.

To define the inertia tensors, one considers infinitesimal rotations and isorotations of the static field.
Introducing angular velocities
\begin{equation}
    a_j = -i\,\Tr(\tau^j A^\dagger \dot{A}), \quad b_j = -i\,\Tr(\tau^j \dot{B} B^\dagger),
\end{equation}
in isospace and physical space respectively, the resulting kinetic contribution to the energy takes the quadratic form
\begin{equation}
    T = \frac{1}{2} a_i U_{ij} a_j + \frac{1}{2} b_i V_{ij} b_j - a_i W_{ij} b_j.
\end{equation}
The matrices $U_{ij}$, $V_{ij}$, and $W_{ij}$ are the isospin, spin, and mixed inertia tensors of the Skyrmion.
Explicitly, these tensors are given by \cite{Manton_2014}
\begin{subequations}
    \begin{align}
        \label{eq: U inertia tensor}
        U_{ij} & = -\int_{\mathbb{R}^3}\textup{d}^3x\, \Tr\!\left( T_i T_j + \frac{1}{4}[L_k,T_i][L_k,T_j] \right), \\
        \label{eq: V inertia tensor}
        V_{ij} & = -\int_{\mathbb{R}^3}\textup{d}^3x\, \varepsilon_{ilm} \varepsilon_{jnp} r^l r^n \Tr\!\left( L_m L_p + \frac{1}{4}[L_k,L_m][L_k,L_p] \right), \\
        \label{eq: W inertia tensor}
        W_{ij} & = \int_{\mathbb{R}^3}\textup{d}^3x\, \varepsilon_{jlm} r^l \Tr\!\left( T_i L_m + \frac{1}{4}[L_k,T_i][L_k,L_m] \right),
    \end{align}
\end{subequations}
where
\begin{equation}
    T_j = \frac{i}{2} U^\dagger[\tau^j,U]
\end{equation}
is an $\mathfrak{su}(2)$-valued current.

The tensor $U_{ij}$ measures the resistance of the Skyrmion to isorotations and depends only on the internal structure of the field.
The tensor $V_{ij}$ characterizes the response to spatial rotations and depends explicitly on the spatial distribution of the energy and currents relative to the centre of mass.
The mixed tensor $W_{ij}$ encodes the coupling between spatial and isorotational motion and vanishes for configurations with sufficient symmetry, such as the spherically symmetric $B=1$ Skyrmion.


\subsection{Electric quadrupole moment}

The intrinsic electric quadrupole tensor is defined by \cite{Wood_2006,Haberichter_2016}
\begin{equation}
\label{eq: Q tensor}
    Q_{ij} = \int_{\mathbb{R}^3} \textup{d}^3x\, \left( 3 r_i r_j - |\mathbf{r}|^2 \delta_{ij} \right) \rho(\mathbf{r}),
\end{equation}
where $\rho(\mathbf{r})$ is the electric charge density.
For isospin-$0$ Skyrmions, the electric charge density is proportional to the baryon density and is given by
\begin{equation}
    \rho(\mathbf{r}) = \tfrac{1}{2}\mathcal{B}^0(\mathbf{r}).
\end{equation}
The quadrupole tensor is traceless by construction and encodes information about the intrinsic deformation of the Skyrmion relative to spherical symmetry.


\subsection{Monopole form factor}

The monopole D-term form factor, which characterizes the internal force distribution of the Skyrmion, is defined by \cite{Martin-Caro_2023,Adam_2024}
\begin{equation}
\label{eq: D-term}
    D = -\frac{2E}{5} \int_{\mathbb{R}^3} \textup{d}^3x\,
    \left( r_i r_j T_{ij} - \frac{1}{3}|\mathbf{r}|^2 T \right),
\end{equation}
where $E$ is the total static energy of the Skyrmion, $T_{ij}$ is the spatial stress tensor, and $T = T_{kk}$ denotes its trace.
The D-term is evaluated in the centre-of-mass frame and provides a quantitative measure of the balance between attractive and repulsive forces inside the soliton.


\subsection{Virial constraint}

An important diagnostic of numerical convergence and physical consistency is provided by the virial (or Derrick) constraint.
For a static field configuration in three spatial dimensions, a necessary condition for stability under uniform rescalings of space is that the total energy be stationary with respect to dilations.
This requirement leads to a virial identity which must be satisfied by any local minimizer of the Skyrme energy.

Concretely, consider a one-parameter family of scaled configurations $U_\lambda(\mathbf{x}) = U(\lambda \mathbf{x})$, with corresponding energy $E(\lambda)$.
For a true static solution, the first derivative of the energy with respect to $\lambda$ must vanish at $\lambda=1$,
\begin{equation}
    \left.\frac{\textup{d}E(\lambda)}{\textup{d}\lambda}\right|_{\lambda=1}=0.
\end{equation}
In field-theoretic terms, this condition is equivalent to the vanishing of the spatial trace of the energy-momentum tensor integrated over space,
\begin{equation}
\label{eq: Virial}
    V = \int_{\mathbb{R}^3} \textup{d}^3x\, T_{kk}.
\end{equation}
Thus, for any static Skyrmion which locally minimizes the energy, one must have $V=0$.

Physically, this constraint expresses the balance between attractive and repulsive contributions to the energy density.
In the Skyrme model, the quadratic sigma-model term, the quartic Skyrme term, and the mass term scale differently under dilations, and a non-trivial solution exists only when these competing effects are in equilibrium.
Locally, the trace $T_{kk}(\mathbf{x})$ need not vanish and typically changes sign within the soliton, but its integral must cancel exactly.

In numerical simulations, deviations of $V$ from zero provide a sensitive measure of how closely a configuration approximates a true solution of the Euler-Lagrange equations.
As such, the virial integral \eqref{eq: Virial} serves as an important consistency check alongside energy convergence and baryon number conservation.


\section{Programme design and implementation in CUDA C}
\label{sec: Program design and implementation in CUDA C}

\begin{figure}[!htp]
  \centering
\begin{tikzpicture}[node distance=2cm]
  \node (cuSkyrmion) [module] {\texttt{cuSkyrmion.cpp}};
\node (interactions) [header, above of=cuSkyrmion] {\texttt{interactions.h}};
\node (skyrmeKernelheader) [header, left of=cuSkyrmion,xshift=-2cm] {\texttt{skyrmeKernel.h}};
\node (settings) [header, above of=skyrmeKernelheader] {\texttt{settings.h}};
\node (skyrmeKernel) [module, below of=skyrmeKernelheader] {\texttt{skyrmeKernel.cu}};
\node (renderKernelheader) [header, right of=cuSkyrmion,xshift=2cm] {\texttt{renderKernel.h}};
\node (renderKernel) [module, below of=renderKernelheader] {\texttt{renderKernel.cu}};
\node (operators) [header, below of=skyrmeKernel] {\texttt{operators.cuh}};
\draw [arrow] (skyrmeKernelheader) -- (cuSkyrmion);
\draw [arrow] (renderKernelheader) -- (cuSkyrmion);
\draw [arrow] (interactions) -- (cuSkyrmion);
\draw [arrow] (skyrmeKernel) -- (skyrmeKernelheader);
\draw [arrow] (renderKernel) -- (renderKernelheader);
\draw [arrow] (operators) -- (skyrmeKernel);
\draw [arrow] (settings) -- (skyrmeKernelheader);
\draw [arrow] (skyrmeKernelheader) -- (interactions);
\end{tikzpicture}
\caption{The \texttt{cuSkyrmion} design. }
\label{fig:modules}
\end{figure}

The code is organized into a small number of modules separating physics evolution, visualization, and user interaction.
The overall structure is illustrated in Fig.~\ref{fig:modules}.
The main translation unit \texttt{cuSkyrmion.cpp} handles programme initialization, GPU memory allocation, time-stepping control, and CUDA--OpenGL interoperability.
Physics kernels are implemented in \texttt{skyrmeKernel.cu} with interfaces defined in \texttt{skyrmeKernel.h}.
Visualization is implemented independently in \texttt{renderKernel.cu} with corresponding declarations in \texttt{renderKernel.h}.
Low-level algebraic operators and device-side helper functions are provided in \texttt{operators.cuh}.
User interaction logic is separated into \texttt{interactions.h}, and finally the settings of the programme are set in \texttt{settings.h} for users to have a single place to set up the programme to their needs.

This separation allows the numerical evolution kernels to remain independent of visualization and UI logic, facilitating extension of the physics implementation without modification of rendering or input handling code.


\subsection{FreeGlut}

The interactive front-end uses FreeGLUT combined with CUDA--OpenGL interoperability.
The application initializes an OpenGL context and registers standard GLUT callback functions for rendering, keyboard input, mouse interaction, window reshape, pop-up menu, and idle updates.
These callbacks primarily trigger GPU-side execution rather than CPU-side rendering.

Rendering is performed entirely on the GPU using a pixel buffer object (PBO) shared between OpenGL and CUDA.
At each frame, the PBO is mapped into CUDA address space using \texttt{cudaGraphicsGLRegisterBuffer} and related mapping calls.
The rendering kernel writes directly into the mapped buffer, after which the buffer is unmapped and displayed using standard OpenGL drawing.
This approach avoids host-device transfers during rendering and ensures that visualization cost scales primarily with GPU memory bandwidth and arithmetic throughput.

The simulation parameters and GPU launch configuration are stored in a single global configuration file \texttt{settings.h}. 
The default computational grid is $N_x \times N_y \times N_z = 151^3$, with stencil radius \texttt{RAD=2}, but other lattice sizes can be chosen in the configuration file \texttt{settings.h}.
This halo width is required by the fourth-order finite difference discretization used for spatial derivatives.


\subsection{Ray tracing}

Visualization is implemented using GPU volume ray tracing through a scalar field derived from the Skyrme configuration.
The baryon density is computed on the simulation lattice and uploaded to a 3D CUDA texture, which is then used for the ray tracing to compute the current visualization of the Skyrmion configuration.

The Skyrme field is stored on the lattice as a normalized four-component field using the \texttt{double4} data type on the device.
From this field the code computes the baryon density and a visualization-oriented floating-point representation suitable for texture upload, which is of type \texttt{float4} and contains also the normalized pion vector. This enables the visualization kernel to directly and locally access the pion data for suitable colouring procedure.
These quantities are computed using dedicated GPU kernels defined in \texttt{skyrmeKernel.cu}.

The visualization volume is stored in a CUDA 3D array and accessed through a texture object.
Texture-based sampling enables hardware interpolation and cache-optimized spatial locality during ray tracing.
The texture is a nomenclature in computer graphics, which in other branches of science is simply known as a linear interpolation function, in this case over a 3-dimensional field.
This makes it possible to query the Skyrmion's baryon density at an arbitrary point in the lattice during the ray tracing, regardless of whether the point exists on the lattice or not. 

Rendering is performed by the kernel \texttt{renderFun} in \texttt{renderKernel.cu}.
For each output pixel, a ray is constructed in world coordinates using an inverse view matrix (i.e.~a rotation matrix from the lattice coordinates to the world coordinates, depending on the current view point) stored in constant memory.
The ray is intersected with the simulation domain bounding box and advanced through the volume using fixed step integration.
At each step the baryon density texture is sampled and converted to colour and opacity using a transfer function controlled by user parameters.
Integration terminates when accumulated opacity exceeds a predefined threshold.

The loop integrating the ray tracing through the rendering volume
computes the RGBA colour 4-vector, with values of each component in
$[0,1]$ as\footnote{In the case of an empty configuration and only in
this case, the maximum baryon density vanishes. In this case, the
\texttt{colour} variable is 4 \texttt{NaN}, but this is then safe
guarded by the built-in CUDA function \texttt{\_\_saturatef} used to
clamp the colour variables to the interval $[0,1]$. When \texttt{NaN}
is passed to \texttt{\_\_saturatef} it returns 0. }
\begin{align}
\texttt{input} &= \left(\frac{\mathcal{B}^0(x)}{\max(\mathcal{B}^0(x))} - \texttt{levelset}\right)\texttt{TRANSFER\_SCALE},\\
\texttt{colour} &= \texttt{input}\left(\texttt{input}\;\hat\phi^1,\texttt{input}\;\hat\phi^2,\texttt{input}\;\hat\phi^3,1\right),
\end{align}
where the maximal baryon density, $\max(\mathcal{B}^0(x))$, is computed in the \texttt{skyrmeKernel.cu} in advance, \texttt{levelset} (e.g.~0.2) is set by the user, \texttt{TRANSFER\_SCALE} is a constant (i.e.~4), 
negative \texttt{input} is rejected (skipped), 
$\hat\phi^a$, $a=1,2,3$ are the three normalized pions,
and finally, the integration measure of the ray tracing is given by a front-to-back blending as
\beq
\d c = 1 - \texttt{input}.
\eeq
The integration is stopped once the alpha channel (the last component of the colour vector, \texttt{colour}, reaches the threshold (i.e.~set to $0.95$) and finally the end result of the integration is multiplied by a \texttt{brightness} constant.

Visualization parameters such as zoom, \texttt{brightness}, and \texttt{levelset} are passed to the kernel each frame, allowing dynamic adjustment without interrupting numerical evolution.


\section{User manual}
\label{sec: User manual}


\subsection{Download, installation and compilation}

The CUDA C code is publicly available on
\begin{verbatim}
https://bitbucket.org/sbgudnason/cuSkyrmion/
\end{verbatim}
for download.
The code is compiled using \texttt{nvcc} and the supplied Makefile.
The primary dependencies are the CUDA toolkit, OpenGL libraries, FreeGLUT and libPNG.
The build system follows the structure of NVIDIA CUDA sample projects and uses \texttt{findgllib.mk} to locate OpenGL and GLUT libraries on the host system.
On a Debian/Ubuntu/Mint Linux system, the installation procedure for the dependencies would be
\begin{verbatim}
sudo apt install nvidia-cuda-toolkit nvidia-cuda-samples libglut-dev libpng-dev
\end{verbatim}
It is assumed that the Linux machine has an NVIDIA graphics card and that the NVIDIA graphics driver is already installed, preferably a recent version.

Compilation is performed using
\begin{verbatim}
make
\end{verbatim}
in the directory of the \texttt{cuSkyrmion} code.
For under-the-hood options and extensions, see Sec.~\ref{sec:under_the_hood}.

On Windows, the cuda-toolkit can be installed from 
\begin{verbatim}
https://developer.nvidia.com/cuda-downloads
\end{verbatim}
whereas the libPNG library and header files (source files) can be downloaded from
\begin{verbatim}
https://gnuwin32.sourceforge.net/packages/libpng.htm
\end{verbatim}

FreeGLUT and libPNG also work for MacOS, but usually such computers do not come with NVIDIA graphics cards, so cannot be used for CUDA software.


\subsection{Running the cuSkyrmion}

At runtime the programme initializes CUDA and OpenGL interoperability and opens an interactive visualization window.

The programme can be run from a terminal
\begin{verbatim}
./cuSkyrmion
\end{verbatim}
which also makes it possible to pass command-line options to the programme at start up.
Alternatively, a programme launcher can be made for the desktop environment on the user's system; an example \texttt{cuSkyrmion.desktop} for GNOME comes with the code.
This file should be copied to the users local GNOME application links by
\begin{verbatim}
desktop-file-install --dir=$HOME/.local/share/applications cuSkyrmion.desktop
\end{verbatim}
Remember to correct the path to \texttt{cuSkyrmion} in the \texttt{.desktop} file before installing it.

Command line arguments are a useful way to start \texttt{cuSkyrmion} with the desired options or configurations, see Sec.~\ref{sec:command_line_arguments}


\subsection{Creating Skyrmions}\label{sec:creating_skyrmions}

Skyrmions can be created at start-up using a configuration file or inserted on-the-fly during run time.

\subsubsection{Configuration file at start-up}\label{sec:config_file}

Starting \texttt{cuSkyrmion} with a configuration file is a precise way of setting up an initial condition for Skyrmion computations.
The format of the configuration file is as follows:
The necessary line for each rational-map Skyrmion to be created, is the line \verb|B=4| where 4 is an example.
The rational-map code only includes $B=1,2,\ldots,9$ -- the reason is the fullerene-type rational map Skyrmion are unstable for $B>7$ when the pion mass is nonvanishing \cite{Battye_2005,Battye_2006}. 
Three optional lines may follow each Skyrmion in arbitrary order, specifying the position \texttt{x}, the orientation \texttt{alpha} and the isospin orientation \texttt{beta}.
The orientations are given as Euler angles in radians with the first being a rotation about the $z$ axis, the second a subsequent rotation about the $y$ axis and the final being a subsequent rotation about the $z$ axis (with the second angle vanishing, the first and the third are equivalent).
An explicit example of a 7-Skyrmion at $x=-1$, $y=0$, $z=0$ rotated by $\pi$ about the $z$ axis, and isorotated by $\pi/2$ about the $y$ axis, is a configuration file:
\begin{verbatim}
B=7
x=(-1,0,0)
alpha=(3.14159,0,0)
beta=(0,1.57079,0)
\end{verbatim}
An example of two $B=4$ Skyrmions is
\begin{verbatim}
B=4
x=(-1.5,0,0)
B=4
x=(1.5,0,0)
alpha=(1.57079,1.57079,-1.57079)
\end{verbatim}
Notice the absence of rotation and isorotations for the first Skyrmion (since they are in the standard orientation \texttt{(0,0,0)}. 
A missing position of a Skyrmion places it at the standard position, i.e.~the origin (\texttt{(0,0,0)}). 
The latter example shows how to insert the twisted-chain $B=8$ Skyrmion -- notice that it is constructed from rational-map constituents and formed by the product ansatz.
Once the user has written the configuration file, it can be loaded from the terminal by
\begin{verbatim}
./cuSkyrmion --configfile config.txt
\end{verbatim}

The rational maps for $B=1,2,\ldots,9$ are those minimizing the energy and given in Ref.~\cite{Houghton_1998}.
The profile functions are fits to the numerically computed profile function (using ordinary differential equations (ODEs)) and they are computed for the pion mass $m=1$ (in Skyrme units) as well as for massless pions, $m=0$.
When \texttt{cuSkyrmion} generates a rational map Skyrmion, it uses a linear interpolation between the two profile functions computed at $m=0$ and $m=1$ if the pion mass parameter is in the range $m\in[0,1]$ and it uses the $m=1$ profile function for $m>1$.
For $m=0,1$ the profile function is exact, but only in the rational map approximation -- the Skyrmion is not a solution to the full equations of motion.

\subsubsection{Inserting Skyrmion on-the-fly}

The rational map Skyrmions, using the same code as described in Sec.~\ref{sec:config_file}, can be inserted at run-time using either the number keys 1 through 9 on the keyboard or the pop-up menu (right click, \texttt{Insert Rational Map} $\to$ \texttt{1-Skyrmion} $\cdots$ \texttt{9-Skyrmion}).
This starts an insertion mode of the programme, where the arrested-Newton flow (or gradient flow) is temporarily stopped and the status bar is asking for the position, orientation and iso-orientation through 6 queries:
\begin{enumerate}
\item Position in the ($x$,$y$)-plane.
\item Position in the ($y$,$z$)-plane.
\item Spatial rotation by varying $\alpha_1$ and $\alpha_2$.
\item Spatial rotation by varying $\alpha_2$ and $\alpha_3$ for fixed $\alpha_1$.
\item Iso-rotation by varying $\beta_1$ and $\beta_2$.
\item Iso-rotation by varying $\beta_2$ and $\beta_3$ for fixed $\beta_1$.
\end{enumerate}
The parameters are adjusted by dragging with the mouse (click, hold and move the mouse) -- while the Skyrmion is being moved or rotated, the status bar in the bottom of the screen displays the current position or angle that is being set.
After each step one must press `\texttt{Enter}' on the keyboard.

The code uses the product ansatz between the existing Skyrmion configuration and the new Skyrmion that is being added in such a way that the would-be outcome is visualized while the user is still deciding where and how to place the new Skyrmion.

If the user is content with the standard settings of the parameter for the new Skyrmion, they may simply press `\texttt{Shift+Enter}'.
This is also possible at any step, `\texttt{Shift+Enter}' will conclude, merge the new Skyrmion with the existing configuration and resume arrested-Newton flow (if the flow mode is not paused).

\paragraph{Sm\"org\aa sbord}

Another way to generate Skyrmion configurations, is to utilize the sm\"or\-g\aa sbord generator, that randomly places $B$ 1-Skyrmions in the configuration space at random orientations (orientation and iso-orientation is equivalent for the 1-Skyrmion).
The Skyrmion number is set using the arrow keys: \texttt{Arrow-Up} increases the baryon number and \texttt{Arrow-Down} decreases it. 
The default value of the baryon number for the sm\"org\aa sbord generator is $B=12$.

To generate the sm\"org\aa sbord, simply press `\texttt{Ctrl+g}' on the keyboard, or right-click to access the pop-up menu and select \texttt{Generate Sm\"org\aa sbord}. 
Notice, that the sm\"org\aa sbord generator distributes the 1-Skyrmions randomly, but no longer apart than the distance set by the \texttt{SMORGAASBORD\_MAXLEN} variable (default value is 2) that can changed in the \texttt{setting.h} file, see Ref.~\ref{sec:under_the_hood} for tuning of the programme.
Notice that the sm\"org\aa sbord generator can be used at any time, inserting $B$ 1-Skyrmions on top of the existing configuration, but the sm\"org\aa sbord generator does not know about the position of the existing Skyrmion -- accidentally (near-)coincident Skyrmions will quickly redistribute themselves under the arrested Newton flow algorithm.


\subsection{Saving figures of Skyrmions}

Figures can be saved to \texttt{.png} format directly from the programme which are screenshots of the current state of the Skyrmion (even in insertion sub-mode).
Press `\texttt{Ctrl+p}' to save a figure file, which is automatically named using a 3-digit sequential numbering scheme as \texttt{snap000.png} etc.
Notice, that the number of the saved image has no correlation with the numbers used for saving binary or text data files.
The figure can also be saved by right-clicking to access the pop-up menu and then selecting `\texttt{Save screenshot}'. 
The white background in the programme is not saved in the figure files; the Skyrmion figure is saved on a transparent background to ease and facilitate figure compositions (using external software, like \texttt{GIMP} etc.). 


\subsection{Saving or exporting your Skyrmions to file}\label{sec:saving}

Field configurations can be written to disk in binary (\texttt{Ctrl+s}) or text (\texttt{Ctrl+t}) format.
This can also be done by right-clicking to access the pop-up menu and selecting `\texttt{Save Skyrmion data}' $\to$ `\texttt{to Binary File}' or `\texttt{to Text File}'.
Saving is performed by copying the field from device to host memory and writing the host buffer.
Filenames are generated automatically using a simple sequential 3-digit numbering scheme as \texttt{skyrmion000.bin} for the binary files and \texttt{skyrmion000.dat} for the text files.
Notice, there is no correlation between the naming of the binary and the text files: the programme simply chooses the next available number.
The user is free to rename and archive (or delete) the saved data using their operating system. 

The binary file format uses about 2.3 times less
harddisk\footnote{The factor of 2.3 comes from the
  implementation of saving 15 decimals of the double variable using
  the C flag '\%.15g'. Including the decimal point, the leading
  digit, the tab character, an estimated 10\% chance of the last digit
  being '0' and hence absent, as well as an estimated 50\% chance of a
  minus sign, the ASCII text format uses $18.4$ bytes per double,
  compared to the 8 bytes in the binary format. The factor is thus
  $18.4/8=2.3$.} space and is convenient for saving a Skyrmion that the user wants to look at, at a later time -- which can also be used as a building block for further Skyrmion building.

The text file format takes up more harddisk space, but is easily imported into packages like \texttt{Mathematica} or \texttt{MATLAB}. 

Loading routines verify grid dimensions before importing field data to ensure consistency with the compiled lattice size.

\subsubsection{The text file format}

The text file format has a header line, i.e.~the first line of the file, which is a tab-separated list of parameters for the current Skyrmion configuration.
The header data contains the variables, from left to right:
\beq
\texttt{xmin}
\quad \texttt{ymin}
\quad \texttt{zmin}
\quad \texttt{XLEN}
\quad \texttt{YLEN}
\quad \texttt{ZLEN}
\quad \texttt{m.x}
\quad \texttt{m.y}
\quad \texttt{m.z}
\quad \texttt{m.w}
\quad \texttt{c6}
\nonumber
\eeq
where the range of lattice coordinates are given by 
$x\in[\texttt{xmin},-\texttt{xmin}]$ (and similarly for $y$, $z$), 
the lattice dimensions are $\texttt{XLEN}\times\texttt{YLEN}\times\texttt{ZLEN}$
and the lattice spacing is $h_x=-\frac{2\,\texttt{xmin}}{\texttt{XLEN}-1}$ (and similarly for $y$, $z$).
Of the five parameters, the first mass parameter, \texttt{m.x}, is the standard pion mass in Skyrme units.
For the other parameters, see Sec.~\ref{sec:under_the_hood}. 

After the header, the Skyrmion data is listed in 4 columns as
\begin{center}
\begin{tabular}{llll}
$\phi_0(0,0,0)$&$\phi_1(0,0,0)$&$\phi_2(0,0,0)$&$\phi_3(0,0,0)$\\
$\phi_0(0,0,1)$&$\phi_1(0,0,1)$&$\phi_2(0,0,1)$&$\phi_3(0,0,1)$\\
\vdots&\vdots&\vdots&\vdots\\
$\phi_0(0,0,\texttt{ZLEN}-1)$&$\phi_1(0,0,\texttt{ZLEN}-1)$&$\phi_2(0,0,\texttt{ZLEN}-1)$&$\phi_3(0,0,\texttt{ZLEN}-1)$\\
$\phi_0(0,1,0)$&$\phi_1(0,1,0)$&$\phi_2(0,1,0)$&$\phi_3(0,1,0)$\\
\vdots&\vdots&\vdots&\vdots
\end{tabular}
\end{center}
where the $z$ axis is looped over first, then the $y$ axis and finally the $x$ axis as the outer loop.

\subsubsection{The binary file format}

The binary format is faster to save in and saves harddisk space, but it not human readable (although it can be read by third-party software, like a custom-made \texttt{Python} script).

\texttt{cuSkyrmion} does not need to know whether the user loads a text file or a binary file when loading a saved Skyrmion configuration on start-up.
The binary format contains an integer set to \texttt{1234} as the first byte of the binary file.
If this byte is different, the programme attempts to load the file as a text file instead.
Then the programme's parameters are stored, so that when loading a Skyrmion it knows the used value of the pion mass etc.
After the parameters, a sentinel is stored, which is \texttt{-123456789.0} of type \texttt{double}.
This is made so that a custom version of the programme can save more parameters to the same file format and can load file formats from the standard version of the programme.
In the case of a custom version of the programme with more parameters, files that do not contain all the new parameters will just set them to their default values. 
Finally, after the sentinel all the pion data are stored directly as in host memory to file, avoiding any resources to order the data.

Specifically the binary format is
\begin{verbatim}
(int) 1234
(double) xmin
(double) ymin
(double) zmin
(int) XLEN
(int) YLEN
(int) ZLEN
(double4) m
(double) c6
(double) -123456789.0
(double4) Skyrmion data array
\end{verbatim}
where \texttt{double4} is an array of four \texttt{double}s. 


\subsection{Keyboard shortcuts and menu items}\label{sec:keyboard_shortcuts}

The programme provides an interactive control interface primarily intended for rapid configuration of initial states, switching between flow algorithms, and adjusting visualization parameters during runtime.
All interaction is handled through GLUT keyboard and mouse callbacks implemented in \texttt{interactions.h}.
The interactive interface is not required for batch or scripted use but provides a convenient method for exploratory numerical studies and debugging.

Flow control is performed through single-key toggles.
Arrested Newton flow, full Newton flow, and gradient flow are enabled or disabled using dedicated keys.
When a flow mode is activated, the simulation advances continuously using the selected evolution scheme until the mode is toggled off.
Switching flow modes does not reinitialize the field configuration.

Field configuration management is performed through insertion and generation commands.
Rational-map Skyrmions of baryon number $B=1,\ldots,9$ can be inserted interactively, see Sec.~\ref{sec:creating_skyrmions}.
In insertion mode the user may specify spatial position and orientation parameters using mouse input before finalizing insertion.
A stochastic initialization routine, the sm\"org\aa sbord generator, is also available which generates ensembles of randomly positioned 1-Skyrmion configurations and combines them using the product ansatz.

Saving operations are performed directly from device memory by copying the field to host memory and writing it to disk in either binary or text format, see Sec.~\ref{sec:saving}.

Visualization parameters can be modified during runtime without interrupting numerical evolution.
These parameters include camera zoom, brightness scaling, and levelset threshold used by the transfer function in the volume renderer.
The default keyboard mappings are summarized below for reference (the keys are case sensitive):

\paragraph{Programme control}
\begin{itemize}
    \item \texttt{Esc} or \texttt{Ctrl+q}: quit
    \item \texttt{Ctrl+Delete}: reset to vacuum [$\phi=(1,0,0,0)$ everywhere]
\end{itemize}

\paragraph{Flow modes}
\begin{itemize}
    \item \texttt{a}: toggle arrested Newton flow (\texttt{flowmode=1})
    \item \texttt{n}: toggle Newton flow (\texttt{flowmode=2})
    \item \texttt{f}: toggle gradient flow (\texttt{flowmode=3})
\end{itemize}

\paragraph{Saving}
\begin{itemize}
    \item \texttt{Ctrl+p}: save screenshot (figure) to file
    \item \texttt{Ctrl+s}: save to binary file
    \item \texttt{Ctrl+t}: save to text file
\end{itemize}

\paragraph{Random initial condition}
\begin{itemize}
    \item \texttt{Ctrl+g}: generate the sm\"org\aa sbord random multi-Skyrmion configuration
    \item $\uparrow$: shift sm\"org\aa sbord baryon number $B$ by $+1$ (default is $B=12$)
    \item $\downarrow$: shift sm\"org\aa sbord baryon number $B$ by $-1$
\end{itemize}

\paragraph{Insert rational-map Skyrmion}
\begin{itemize}
    \item \texttt{1--9}: choose baryon number $B$ for insertion and enter insertion mode
    \item \texttt{Enter}: advance insertion sub-mode (position/orientation steps)
    \item \texttt{Shift+Enter}: accept default insertion parameters immediately (can be used at any step during insertion sub-mode)
\end{itemize}

\paragraph{Model parameter}
\begin{itemize}
    \item \texttt{m} / \texttt{M}: decrease/increase the pion mass parameter \texttt{parms.m.x} in small steps $-/+0.01$
\end{itemize}

\paragraph{Extended model parameters (see Sec.~\ref{sec:under_the_hood})}
\begin{itemize}
    \item \texttt{j} / \texttt{J}: decrease/increase the pion mass parameter \texttt{parms.m.y} in small steps $-/+0.01$
    \item \texttt{k} / \texttt{K}: decrease/increase the pion mass parameter \texttt{parms.m.y} in small steps $-/+0.01$
    \item \texttt{l} / \texttt{L}: decrease/increase the pion mass parameter \texttt{parms.m.y} in small steps $-/+0.01$
    \item \texttt{h} / \texttt{H}: decrease/increase the BPS Skyrme term coefficient \texttt{parms.c6} in small steps $-/+0.01$
\end{itemize}
These key strokes are only enabled when the corresponding extension is enabled at compile time, see Sec.~\ref{sec:under_the_hood}.

\paragraph{Visualization parameters}
\begin{itemize}
    \item \texttt{z} / \texttt{Z}: adjust zoom (decrease/increase)
    \item \texttt{[} / \texttt{]}: adjust brightness (decrease/increase)
    \item \texttt{,} / \texttt{.}: adjust the level-set threshold (decrease/increase)
\end{itemize}

\paragraph{Physical observables}
\begin{itemize}
    \item \texttt{R}: print RMS radius $R_{\textup{Sk}}$ \eqref{eq: RMS radius} to terminal
    \item \texttt{U}: print isospin inertia tensor $U_{ij}$ \eqref{eq: U inertia tensor} to terminal
    \item \texttt{V}: print spin inertia tensor $V_{ij}$ \eqref{eq: V inertia tensor} to terminal
    \item \texttt{W}: print mixed inertia tensor $W_{ij}$ \eqref{eq: W inertia tensor} to terminal
    \item \texttt{Q}: print electric quadrupole tensor $Q_{ij}$ \eqref{eq: Q tensor} to terminal
    \item \texttt{D}: print D-term $D(0)$ \eqref{eq: D-term} to terminal
    \item \texttt{T}: print virial constraint $V$ \eqref{eq: Virial} to terminal
    \item \texttt{P}: print model parameters in physical units to terminal (this requires that \texttt{FPI} and \texttt{ESKYRME} are set in \texttt{settings.h}, see Sec.~\ref{sec:under_the_hood})
\end{itemize}

A context menu providing equivalent functionality is available through the GLUT right-click menu.
The menu provides rapid access to flow mode selection, saving operations, and rational-map insertion shortcuts.
The menu implementation directly invokes the same control functions as the keyboard interface to ensure consistent behaviour between input methods.


\subsection{Command line arguments}\label{sec:command_line_arguments}

Configuration files allow scripted construction of multi-Skyrmion initial states, see Sec.~\ref{sec:config_file}.
The command line arguments supported by \texttt{cuSkyrmion} are summarized as follows:
\paragraph{Generate or load Skyrmions on start-up}
\begin{itemize}
\item \texttt{-c config.txt} or \texttt{--configfile config.txt} reads the configuration file \texttt{config.txt} for creating Skyrmions at start-up, see Sec.~\ref{sec:config_file}
\item \texttt{-d file.bin} or \texttt{--datafile file.bin} reads a \texttt{.bin} binary file or a \texttt{.dat} text file with a previously saved Skyrmion configuration
\item \texttt{-g B} or \texttt{--smorgaasbord B} generates \texttt{B} Skyrmions using the sm\"org\aa sbord generator, where \texttt{B} is a positive integer
\end{itemize}
\paragraph{Turn on flow modes on start-up}
\begin{itemize}
\item \texttt{-a} or \texttt{--anewtonflow} turns on arrested Newton flow on start-up
\item \texttt{-n} or \texttt{--newtonflow} turns on Newton flow on start-up
\item \texttt{-f} or \texttt{--gradientflow} turns on gradient flow on start-up
\end{itemize}
\paragraph{Set model parameters on start-up}
\begin{itemize}
\item \texttt{-m X} or \texttt{--pionmass X} sets the pion mass to \texttt{X} which is a positive floating point number
\item \texttt{-j X} or \texttt{--modifiedpion X} sets the modified pion mass \texttt{m.y} to \texttt{X} which is a positive floating point number
\item \texttt{-k X} or \texttt{--looselybound X} sets the loosely bound potential mass \texttt{m.z} to \texttt{X} which is a positive floating point number
\item \texttt{-l X} or \texttt{--lightlybound X} sets the lightly bound potential mass \texttt{m.w} to \texttt{X} which is a positive floating point number
\item \texttt{-h X} or \texttt{--bpskyrme X} sets the BPS Skyrme term coefficient \texttt{c6} to \texttt{X} which is a positive floating point number
\end{itemize}
The order of the command line arguments can be arbitrary.
All values of model parameters should be given in Skyrme units, i.e.~dimensionless units.
For the parameters corresponding to extensions (\texttt{-j}, \texttt{-k}, \texttt{-l} and \texttt{-h}), see Sec.~\ref{sec:under_the_hood}. 

The following alternative variations of passing command line arguments are equivalent:
\begin{verbatim}
./cuSkyrmion -c config.txt
./cuSkyrmion --configfile config.txt
./cuSkyrmion configfile config.txt
./cuSkyrmion --configfile=config.txt
./cuSkyrmion configfile=config.txt
\end{verbatim}
and equivalently for the other parameters
Notice that the abbreviated version, \texttt{-c}, must be with the dash and cannot be followed by the equal sign (\texttt{=}). 

An example of combined command line arguments is:
\begin{verbatim}
./cuSkyrmion -c config.txt -m 0.5 -a
\end{verbatim}
which loads the configuration file \texttt{config.txt}, generates the Skyrmions specified in that file, sets the pion mass parameter to \texttt{0.5} and starts the arrested Newton flow algorithm immediately.


\subsection{Under the hood -- activating extensions and modifying them}\label{sec:under_the_hood}

Compile-time macros control optional physical terms and numerical as well as programme and visualization options.
The defined macros in \texttt{settings.h} use the \texttt{C/C++} precompiler to insert the chosen options into the code at every occurrence and is an easy and systematic way of modifying the code at compile time.
The need for modifying the code at compile time, is to optimize performance, so that unnecessary computations are simply not performed, if the extension at hand is not needed or wanted.

We will introduce the \texttt{settings.h} file here, explaining how to modify the code to better match the user's hardware (GPU) and physical model requirements.
The first section of \texttt{settings.h} sets the lattice size dimension, the lattice size (in Skyrme units), the window size (in pixels), the GPU block size and the number of computations done between every frame rendering.
Three predefined settings are given in \texttt{settings.h}; uncomment the one most suitable for your system and requirements and the recompile with '\verb|make clean && make|'.

The macros for the large sized lattice suitable for large and recent GPUs are
\begin{verbatim}
#define XLEN 151
#define YLEN 151
#define ZLEN 151
#define DEFAULT_LATTICE_SIZE 8.
#define WINDOW_WIDTH 1024
#define WINDOW_HEIGHT 1024
#define GPU_BLOCKSIZE_X 32
#define GPU_BLOCKSIZE_Y 32
#define ITERS_PER_RENDER 10
\end{verbatim}
Notice the macros do not define the type, but if there is no '.' at the end, the type is a (positive) integer, whereas '8.' is a floating point type (a decimal number).

The next section of \texttt{settings.h} sets the pion mass' default value:
\begin{verbatim}
#define DEFAULT_PION_MASS 1.
\end{verbatim}
This must be a non-negative floating point number.

The next section allows for activation of physical units (in MeV and fm) by setting $F_\pi$ and $e$, the Skyrme coupling constant:
\begin{verbatim}
//#define FPI 129.
//#define ESKYRME 5.45
\end{verbatim}
where it is understood that \texttt{FPI} is in MeV.

Then the \texttt{COURANT} variable can be set, which controls the flow speed, where the fictive time step size is set to be
\beq
h_t = \texttt{COURANT}\times h_x^4.
\eeq
If the lattice is course or the BPS-Skyrme term extension has been turned on, one may need to lower the value of this constant:
\begin{verbatim}
#define COURANT 0.1
\end{verbatim}

Then the visualization options can be modified if needed:
\begin{verbatim}
#define DEFAULT_ZOOM 5.5f
#define DEFAULT_BRIGHTNESS 5.0f
#define DEFAULT_LEVELSET 0.2f
#define STATUSBAR_TIMEOUT 5000
\end{verbatim}
which correspond to the default values of zoom, brightness, levelset of the baryon charge density (between 0 and 1) and finally the timeout of status bar messages in milliseconds. 

The next section of \texttt{settings.h} allows to enable the extension models which are variant models of the Skyrme model:
\begin{verbatim}
//#define MODIFIED_PION_POTENTIAL
//#define LOOSELY_BOUND_POTENTIAL
//#define LIGHTLY_BOUND_POTENTIAL
//#define BPS_SKYRME_TERM
\end{verbatim}
By default they are commented out (\texttt{//}).
They may also be activated directly from the command line by using
\begin{verbatim}
make clean && make EXTRA_CCFLAGS=-DLIGHTLY_BOUND_POTENTIAL
\end{verbatim}
as an example.

Finally, the default values of the extensions are set:
\begin{verbatim}
#define DEFAULT_MODIFIED_PION_MASS 0.
#define DEFAULT_LOOSELY_BOUND_MASS 0.
#define DEFAULT_LIGHTLY_BOUND_MASS 0.
#define DEFAULT_C6 1.
\end{verbatim}

The mass parameters in Skyrme unit (dimensionless units) correspond to the four potentials
\begin{align}
2m_x^2(1-\phi_0), \quad
m_y^2(1-\phi_0^2),\quad
m_z^2(1-\phi_0)^2,\quad
\frac12m_w^2(1-\phi_0)^4,
\end{align}
which correspond the standard pion mass (default potential) with pion mass $m_x$ \cite{Adkins_1984}, the modified pion mass potential with pion mass $m_y$ \cite{Gudnason:2014nba}, the loosely bound potential with mass $m_z$ \cite{Gudnason:2016mms}, and the lightly bound potential with mass $m_w$ \cite{Harland:2013rxa,Gillard:2015eia}.
The naming of the masses correspond to their variables in the programme as they are stored as a \texttt{double4} object and are accessed as \texttt{m.x}, \texttt{m.y}, \texttt{m.z} and \texttt{m.w}, respectively.
Notice that the three first potentials are not linearly independent, either the modified pion mass or the loosely bound potential can be eliminated by changing also the pion mass (although eliminating the former corresponds to a case where the parameter $m_z^2$ may be negative, see Ref.~\cite{Gudnason:2016mms}, which the programme does not allow for).

The coefficient, $c_6$ (with variable name \texttt{c6}) is the coefficient, in Skyrme units, of the optional BPS Skyrme term \cite{Adam:2010fg,Adam:2010ds}
\beq
\mathcal{L}_6 = c_6 4\pi^4\big(\mathcal{B}^0\big)^2,
\eeq
which is sextic in spatial derivatives.


\subsection{Troubleshooting}

\paragraph{The code is not compiling}
This may be a compilation error or a linker error.
Check that:
\begin{itemize}
\item The CUDA compiler and required libraries are installed; ensure the development libraries are installed, so the needed header (\texttt{.h}) files can be accessed at compile-time. 
\item No macro is defined more than once in \texttt{settings.h} file.
\end{itemize}

\paragraph{Black window on programme start-up}
Check that:
\begin{itemize}
\item The required OpenGL extensions are installed and can be found by the linker.
\end{itemize}

\paragraph{My stored datafile does not load}
\begin{itemize}
\item Make sure that the lattice size is exactly the same in the currently compiled version of the programme as in the saved datafile.
\end{itemize}

\paragraph{The Skyrmion does not converge}
The residue variable \texttt{delta} does not converge to a small number during flow.
Check that:
\begin{itemize}
\item The flow must be arrested Newton flow (press `\texttt{a}') or gradient flow (press `\texttt{f}') for convergence. Newton flow does not converge to a minimum of the energy functional. 
\item If potential parameters are large or the BPS-Skyrme term extension is turned on, or if the lattice spacing is too course, the variable \texttt{COURANT} may be too large. Try smaller values of \texttt{COURANT} until convergence is achieved.
\end{itemize}

\paragraph{I deleted my configuration, but the Skyrmion is still showing}
The visualization buffer and computation memory are not the same.
\begin{itemize}
\item Start a flow mode, e.g.~press `\texttt{a}' on the keyboard to update the window.
\end{itemize}


\section{Neat examples}
\label{sec: Neat examples}

Typical usage consists of generating or loading an initial configuration, evolving it using arrested Newton or gradient flow, visualizing the evolving configuration in real time, observing that the computational residue \texttt{delta} becomes small (say below $10^{-3}$ or $10^{-5}$) and saving the final state to datafile and print a screenshot from a suitable angle.


\subsection{\texorpdfstring{$B=3$}{B=3} Skyrmion}
\label{sec:B=3example}

The following usage is an example workflow for creating a $B=3$ Skyrmion, with default dimensionless pion mass $m=1$, on a cubic grid with lattice size $N=151^3$:
\begin{enumerate}
    \item Compile and run: \texttt{make} and then \texttt{./cuSkyrmion} (or run \texttt{cuSkyrmion.exe} in Windows).
    \item Insert $B=3$ rational map Skyrmion by pressing `\texttt{3}' followed by `\texttt{Shift}+\texttt{Enter}'.
    \item Run arrested Newton flow by pressing `\texttt{a}' and run until convergence.
    \item The relaxed baryon number and energy should, respectively, read approximately $B=2.998$ and $E=470.13$.
    \item The inertia tensors $U,V,W$ can be obtained by pressing `\texttt{U}', `\texttt{V}' and `\texttt{W}'. They should be diagonal and isotropic ($U_{ij}=u\delta_{ij}$, $V_{ij}=v\delta_{ij}$, $W_{ij}=w\delta_{ij}$), and approximately be
    \begin{equation}
       u=124.16, \quad v=403.13, \quad w=-85.27.
    \end{equation}
    \item The RMS radius $R$ can be obtained by pressing `\texttt{R}', and is roughly $R=1.23$.
    \item The virial constraint is obtained by pressing `\texttt{T}', and should be $\textup{Tr}(S) = 0.00$.
    \item The isospin-0 electric quadrupole tensor $Q$ is obtained by pressing `\texttt{Q}'. It should be diagonal, and zero in this case, with $\textup{Tr}(Q) = 0.00$.
    \item Finally, the D-term can be obtained by pressing `\texttt{D}' and should be approximately $D(0) = -41.55$.
    \item Save a screenshot of the $B=3$ configuration to a PNG file by pressing `\texttt{Ctrl}+\texttt{p}'.
\end{enumerate}

\begin{figure}
    \centering
     \begin{minipage}[b]{0.45\linewidth}
     \begin{verbatim}
B=3
    \end{verbatim}
    \end{minipage}
   \begin{minipage}[b]{0.5\linewidth}
    \includegraphics[width=0.5\linewidth]{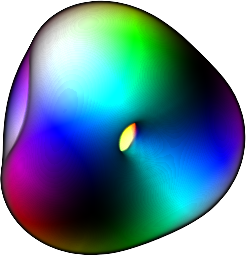}
    \end{minipage}
    \caption{The massive $B=3$ Skyrmion obtained from the rational map
      ansatz on an $N=151^3$ grid with pion mass parameter $m=1$.
      The convergence tolerance used is $\delta<10^{-5}$.
    }
    \label{fig:B3}
\end{figure}
The screenshot from the programme is shown in Fig.~\ref{fig:B3}.

\subsection{\texorpdfstring{$B=8_a$}{B=8a} Skyrmion}
\label{sec:B=8a}

\begin{figure}
    \centering
     \begin{minipage}[b]{0.45\linewidth}
     \begin{verbatim}
B=4
x=(-1.5,0,0)
B=4
x=(1.5,0,0)
alpha=(1.57079,1.57079,-1.57079)
    \end{verbatim}
    \end{minipage}
   \begin{minipage}[b]{0.5\linewidth}
    \includegraphics[width=0.8\linewidth]{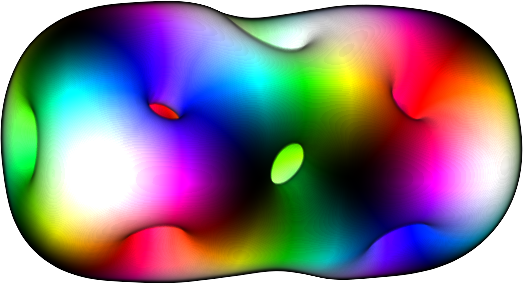}
    \end{minipage}
    \caption{The massive $B=8_a$ Skyrmion obtained from the config
      file on the left in the figure on an $N=151^3$ grid with pion
      mass parameter $m=1$.
      The convergence tolerance used is $\delta<10^{-5}$.
    }
    \label{fig:B8a}
\end{figure}

For this $B=8_a$ example, we obtain the following observables, see Fig.~\ref{fig:B8a},
$B=7.997$, $E=1212.37$, $R=2.02$, $\Tr(Q)=0$, $\Tr(S)=0$, $D(0)=-209.04$.
The subscript refers to the solutions in Ref.~\cite{Gudnason_2022}.
The tensors are given by
\begin{align}
U &= 
\begin{pmatrix}
293.39 & -3.18 & 0\\ 
-3.18 & 297.06 & 0\\ 
0 & 0 & 326.30
\end{pmatrix},\quad&
V &= 
\begin{pmatrix}
1388.92 & 0.03 & 0.02\\
0.03 & 4055.67 & 0\\
0.02 & 0 & 4055.67
\end{pmatrix},\non
W &= 0,\quad&
Q &= \diag\left(16.78, -8.39, -8.39\right).
\end{align}

\subsection{\texorpdfstring{$B=12_a$}{B=12a} Skyrmion}
\label{sec:B=12a}

\begin{figure}
    \centering
     \begin{minipage}[b]{0.45\linewidth}
     \begin{verbatim}
B=4
x=(-3,0,0)
B=4
alpha=(1.57079,1.57079,-1.57079)
B=4
x=(3,0,0)
    \end{verbatim}
    \end{minipage}
   \begin{minipage}[b]{0.5\linewidth}
    \includegraphics[width=0.8\linewidth]{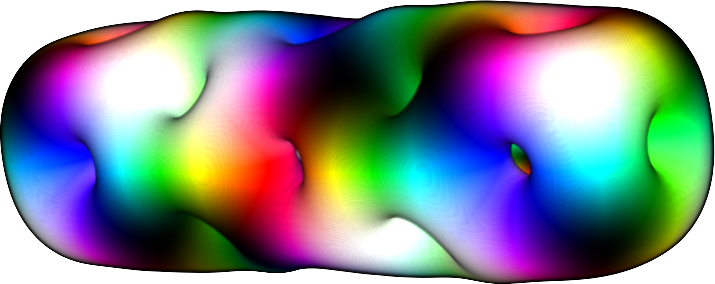}
    \end{minipage}
    \caption{The massive $B=12_a$ Skyrmion obtained from the config
      file on the left in the figure on an $N=151^3$ grid with pion
      mass parameter $m=1$.
      The convergence tolerance used is $\delta<10^{-5}$.
    }
    \label{fig:B12a}
\end{figure}

For this $B=12_a$ example, which is the ground state for $m=1$, we obtain the following observables, see Fig.~\ref{fig:B12a},
$B=11.995$, $E=1810.72$, $R=2.79$, $\Tr(Q)=0$, $\Tr(S)=0$,
$D(0)=-431.00$.
The subscript refers to the solutions in Ref.~\cite{Gudnason_2022}.
The tensors are
\begin{align}
U &= 
\begin{pmatrix}
441.46 & -4.25 & 0\\
-4.25 & 446.37 & 0\\
0 & 0 & 474.18
\end{pmatrix},\quad&
V &= 
\diag\left(2095.46, 12854.75,12854.75\right),\non
W &= 0,\quad&
Q &= \diag\left(69.36, -34.68, -34.68\right).
\end{align}

\subsection{\texorpdfstring{$B=12_b$}{B=12b} Skyrmion}
\label{sec:B=12b}

\begin{figure}
    \centering
     \begin{minipage}[b]{0.45\linewidth}
     \begin{verbatim}
B=7
x=(-2,0,0)
B=5
x=(2,0,0)
alpha=(0,0,-1.57079)
    \end{verbatim}
    \end{minipage}
   \begin{minipage}[b]{0.5\linewidth}
    \includegraphics[width=0.8\linewidth]{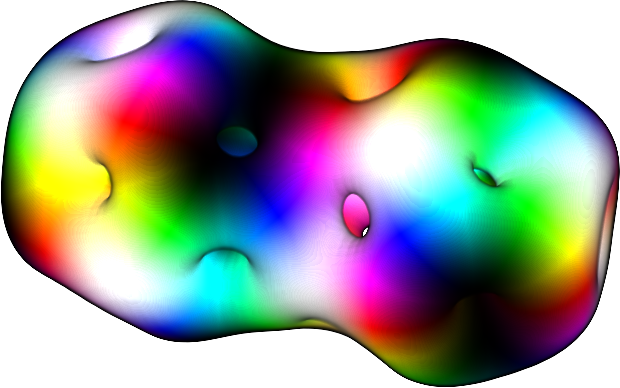}
    \end{minipage}
    \caption{The massive $B=12_b$ Skyrmion obtained from the config
      file on the left in the figure on an $N=151^3$ grid with pion
      mass parameter $m=1$.
      The convergence tolerance used is $\delta<10^{-5}$.
    }
    \label{fig:B12b}
\end{figure}

For this $B=12_b$ example, we obtain the following observables, see Fig.~\ref{fig:B12b},
$B=11.995$, $E=1811.35$, $R=2.37$, $\Tr(Q)=0$, $\Tr(S)=0$, $D(0)=-351.52$.
The subscript refers to the solutions in Ref.~\cite{Gudnason_2022}.
The tensors are
\begin{align}
U &= 
\begin{pmatrix}
444.26 & -1.98 & 2.09\\
-1.98 & 438.55 & 6.63\\
2.09 & 6.63 & 437.91
\end{pmatrix},\quad&
V &= 
\begin{pmatrix}
3289.58 & 553.78 & 643.82\\
553.78 & 8280.52 & -69.37\\
643.82 & -69.37 & 8262.05
\end{pmatrix},\non
W &= 
\begin{pmatrix}
20.67 & -2.25 & -2.65\\
65.80 & -7.22 & -8.38\\
-69.09 & 7.60 & 8.83
\end{pmatrix},\quad&
Q &= 
\begin{pmatrix}
32.56 & -5.43 & -6.31\\
-5.43 & -16.37 & 0.68\\
-6.31 & 0.68 & -16.19
\end{pmatrix}.
\end{align}

\subsection{\texorpdfstring{$B=14_a$}{B=14a} Skyrmion}
\label{sec:B=14a}

\begin{figure}
    \centering
     \begin{minipage}[b]{0.45\linewidth}
     \begin{verbatim}
B=7
x=(-2,0,0)
B=7
x=(2,0,0)
alpha=(1.57079,1.57079,-1.57079)
    \end{verbatim}
    \end{minipage}
   \begin{minipage}[b]{0.5\linewidth}
    \includegraphics[width=0.8\linewidth]{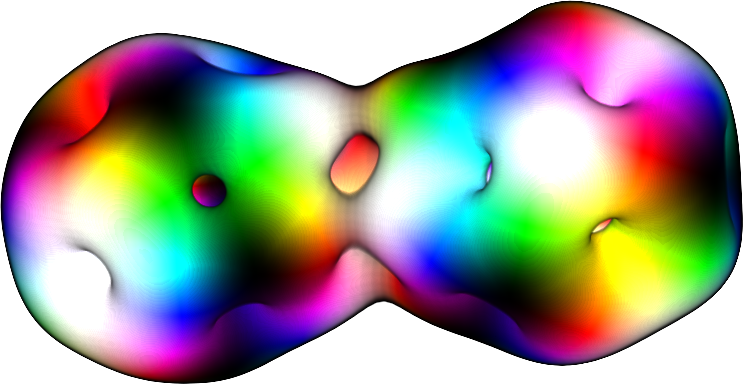}
    \end{minipage}
    \caption{The massive $B=14_a$ Skyrmion obtained from the config
      file on the left in the figure on an $N=151^3$ grid with pion
      mass parameter $m=1$.
      The convergence tolerance used is $\delta<10^{-5}$.
    }
    \label{fig:B14a}
\end{figure}

For this $B=14_a$ example, which is the ground state for $m=1$, we obtain the following observables, see Fig.~\ref{fig:B14a},
$B=13.994$, $E=2106.67$, $R=2.78$, $\Tr(Q)=0$, $\Tr(S)=0$, $D(0)=-526.52$.
The subscript refers to the solutions in Ref.~\cite{Gudnason_2022}.
The tensors are
\begin{align}
U &= 
\begin{pmatrix}
514.52 & -0.23 & 3.55\\
-0.23 & 511.12 & 0.37\\
3.55 & 0.37 & 513.94
\end{pmatrix},\quad&
V &= 
\begin{pmatrix}
3689.58 & 295.88 & 594.92\\
295.88 & 14165.18 & -17.02\\
594.92 & -17.02 & 14139.63
\end{pmatrix},\non
W &= 
\begin{pmatrix}
-0.07 & -0.76 & -0.58\\
0.13 & 1.79 & 1.39\\
0.05 & 0.81 & 0.63
\end{pmatrix},\quad&
Q &= 
\begin{pmatrix}
68.86 & -2.92 & -5.87\\
-2.92 & -34.56 & 0.17\\
-5.87 & 0.17 & -34.31
\end{pmatrix}.
\end{align}

\subsection{\texorpdfstring{$B=16_c$}{B=16c} Skyrmion}
\label{sec:B=16c}

\begin{figure}
    \centering
     \begin{minipage}[b]{0.45\linewidth}
     \begin{verbatim}
B=3
x=(-2,0,0)
B=7
x=(0,0,0)
alpha=(1.57079,1.57079,-1.57079)
B=3
x=(0,2,0)
alpha=(1.57079,0,0)
B=3
x=(2,0,0)
    \end{verbatim}
    \end{minipage}
   \begin{minipage}[b]{0.5\linewidth}
    \includegraphics[width=0.8\linewidth]{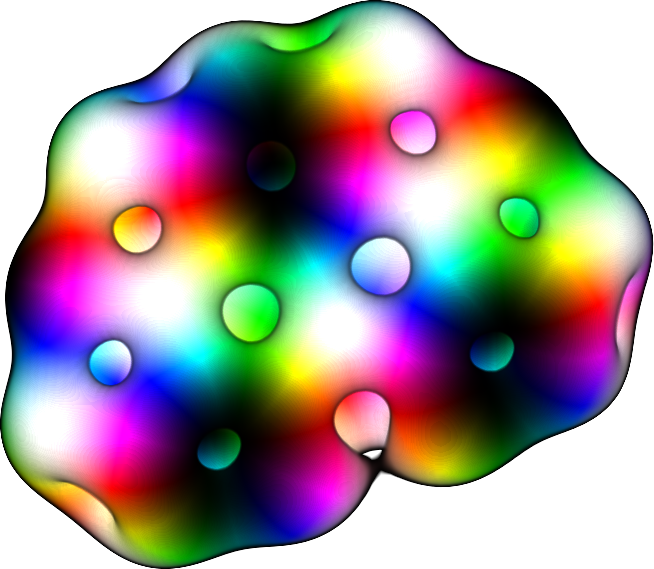}
    \end{minipage}
    \caption{The massive $B=16_c$ Skyrmion obtained from the config
      file on the left in the figure on an $N=151^3$ grid with pion
      mass parameter $m=1$.
      The convergence tolerance used is $\delta<10^{-5}$.
    }
    \label{fig:B16c}
\end{figure}

For this $B=16_c$ example, we obtain the following observables, see Fig.~\ref{fig:B16c},
$B=15.994$, $E=2408.02$, $R=2.62$, $\Tr(Q)=0$, $\Tr(S)=0$, $D(0)=-738.76$. 
The subscript refers to the solutions in Ref.~\cite{Gudnason_2022}.
The tensors are
\begin{align}
U &= 
\begin{pmatrix}
597.70 & -2.23 & -0.52\\
-2.23 & 593.85 & 0.95\\
-0.52 & 0.95 & 602.59
\end{pmatrix},\quad&
V &= 
\begin{pmatrix}
7849.22 & -2152.82 & -98.82\\
-2152.82 & 12950.85 & -1874.63\\
-98.82 & -1874.63 & 11530.65
\end{pmatrix},\non
W &= 
\begin{pmatrix}
42.41 & 55.04 & 27.86\\
-44.16 & 17.03 & 30.24\\
50.35 & -83.90 & -3.81
\end{pmatrix},\quad&
Q &= 
\begin{pmatrix}
29.22 & 21.43 & 0.72\\
21.43 & -21.88 & 19.22\\
0.72 & 19.22 & -7.33
\end{pmatrix}.
\end{align}

\section{Benchmark and scaling}
\label{sec:benchmark}

\begin{table}
\begin{center}
\caption{The run-time in seconds of three differently sized configurations of \texttt{settings.h} on different GPUs. }
\label{tab:scaling}
\begin{tabular}{lrrrr}
& CUDA cores & LARGE SIZE & MEDIUM SIZE & SMALL SIZE\\
GTX 1650 & 1024 & 1438.78 & 869.93 & 178\\
RTX 4090 & 16384 & 116.19 & 72.23 & 18.76\\
RTX 5090D & 21760 & 81.77 & 51.08 & 18.45
\end{tabular}
\end{center}
\end{table}

\begin{figure}[!htp]
\centering
\includegraphics[width=0.5\linewidth]{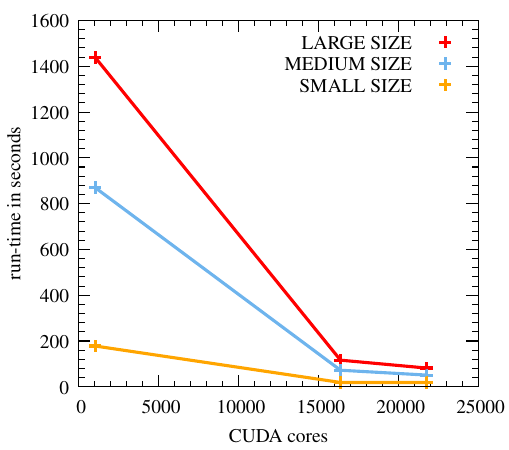}
\caption{Scaling: run-time in seconds of three different configurations versus the number of CUDA cores.}
\label{fig:scaling}
\end{figure}

We compute a sample configuration using \verb|B=4| as the config file on the large, medium and small sized configurations in \texttt{settings.h} and the result is shown in Tab.~\ref{tab:scaling} and Fig.~\ref{fig:scaling}.
The specs of the three different sized configurations are shown in Tab.~\ref{tab:specs}.
The small configuration does not scale well, since it cannot fully utilize the power of the large GPUs.

\begin{table}
\begin{center}
\caption{Specs for three differently sized configurations in \texttt{settings.h}.}
\label{tab:specs}
\begin{tabular}{lrrr}
& LARGE SIZE & MEDIUM SIZE & SMALL SIZE\\
XLEN & 151 & 127 & 65\\
YLEN & 151 & 127 & 65\\
ZLEN & 151 & 127 & 65\\
DEFAULT\_LATTICE\_SIZE & 8. & 7. & 5.\\
WINDOW\_WIDTH & 1024 & 768 & 512\\
WINDOW\_HEIGHT & 1024 & 768 & 512\\
GPU\_BLOCKSIZE\_X & 32 & 32 & 32\\
GPU\_BLOCKSIZE\_Y & 32 & 16 & 8\\
ITERS\_PER\_RENDER & 10 & 5 & 3
\end{tabular}
\end{center}
\end{table}

Since this is the first GPU/CUDA-based Skyrmion software with continuous visualization, we have no previous software to compare the performance to.


\section{The Python port: \texttt{skyrmion\_solver}}
\label{sec: The Python port: skyrmion_solver}

The \texttt{Python} port of \texttt{cuSkyrmion} is \texttt{skyrmion\_solver}.
It is not intended as a literal line-by-line translation of the \texttt{CUDA C} code, but rather as a reimplementation of the same GPU-native numerical philosophy within the modular architecture developed for \texttt{soliton\_solver} \cite{soliton_solver}.
The package is distributed in Python via PyPI (\href{https://pypi.org/project/skyrmion-solver/}{https://pypi.org/project/skyrmion-solver/}) and can, alternatively, be downloaded directly from the public github repository (\href{https://github.com/Paulnleask/skyrmion_solver}{https://github.com/Paulnleask/skyrmion\_solver}).
In particular, the port preserves the structured-grid finite-difference workflow, explicit GPU time-stepping, and GPU-resident rendering pipeline of \texttt{cuSkyrmion}, while reorganizing the code into a reusable framework in which the shared numerical engine is separated from the theory-specific physics.

The resulting software is a modular \texttt{Python} package for three-dimensional Skyrme-type field theories.
It retains the ability to simulate and visualize Skyrmion configurations in real time, but extends the original scope of \texttt{cuSkyrmion} by allowing several related Skyrme models to be implemented within a common execution model.
This makes it possible to treat not only the standard pion-only Skyrme model, but also variants and extensions involving symmetry breaking, vector mesons, and Coulomb backreaction, without rewriting the solver core for each case.

On the other hand, \texttt{skyrmion\_solver} does not come with the same main programme module as \texttt{cuSkyrmion} with an all-in-one file format and insertion modes that can be used on-the-fly to create Skyrmions visually during computations and during run-time.


\subsection{Software architecture}

The package is organized around a strict separation between reusable numerical infrastructure and theory-specific modules.
At the repository level this appears as a \texttt{core/} layer, a \texttt{theories/} layer, a \texttt{visualization/} layer, and an \texttt{examples/} layer.

The \texttt{core/} layer provides the shared numerical backend.
This includes parameter resolution and packing, flattened indexing helpers, finite-difference derivative operators, CUDA launch utilities, reduction kernels, time integration routines, and the \texttt{Simulation} driver that coordinates device memory allocation, stepping, observables, and rendering.
The purpose of the \texttt{core/} layer is to remain theory agnostic: it contains no model-specific Lagrangian or field content.

The \texttt{theories/} layer contains the actual Skyrme-type models.
Each theory module provides its own field content, parameter set, CUDA kernels for local densities and gradients, initialization routines, theory-specific observables, and optional helper functions for rendering and output.
A runtime theory registry stores the metadata needed to load a selected model and inject it into the common simulation workflow.
This is the main architectural difference from \texttt{cuSkyrmion}, which was centred on a single Skyrme code path and compile-time options.
In \texttt{skyrmion\_solver}, the same numerical engine can be combined at runtime with different Skyrme theories without changing the shared backend.

The \texttt{visualization/} layer provides the GPU-resident rendering backend.
This layer is also reusable across theories, since it consumes device-side scalar and vector volumes rather than making assumptions about a specific model.
Finally, the \texttt{examples/} layer contains runnable entry points illustrating the standard workflow for the built-in theories.


\subsection{Software functionalities}

At the level of user-visible functionality, \texttt{skyrmion\_solver} provides the following common capabilities across the built-in theories:
\begin{itemize}
    \item GPU-native finite-difference simulation of non-linear field theories on structured three-dimensional lattices;
    \item explicit time integration, including fourth-order Runge--Kutta evolution;
    \item arrested Newton flow for rapid relaxation to static or metastable multi-Skyrmion configurations;
    \item initialization routines based on hedgehog, rational-map, product, and stochastic multi-Skyrmion ans\"atze;
    \item evaluation of physically relevant observables such as the baryon number, centre of mass, RMS radius, moments of inertia, quadrupole tensor, and monopole D-term;
    \item real-time CUDA--OpenGL volume rendering of energy density and field structure;
    \item export of field configurations and derived quantities for offline analysis.
\end{itemize}

As in \texttt{cuSkyrmion}, the principal design goal is to keep the expensive parts of the workflow on the GPU.
Field arrays, work buffers, derivative buffers, reduction buffers, and rendering volumes remain device resident during the main loop, with host-side interaction largely limited to parameter management, launch coordination, and small reductions over blockwise partial results.
This is particularly important for the three-dimensional Skyrme problem, where the field volumes are large and repeated host--device transfers would become prohibitively expensive.

\subsubsection{Software functionality \texorpdfstring{\emph{not}
    implemented in \texttt{skyrmion\_solver}}{not implemented in
    skyrmion\_solver}}

The following features of the main module of \texttt{cuSkyrmion} are
\emph{not} currently implemented in \texttt{skyrmion\_solver}:
\begin{itemize}
\item Run-time interactive insertion mode with the ability to choose
  the position, orientation and iso-orientation interactively with the
  mouse.
\item Run-time stochastic Skyrmion generation on top of the current
  Skyrmion configuration (the sm\"org\aa sbord generator) [in
    \texttt{cuSkyrmion} the sm\"org\aa sborg generator can also be run
    at program-start using the command-line argument \texttt{-g},
    which is a feature implemented in \texttt{skyrmion\_solver}].
\item Run-time control of the pion mass and other potential
  parameters, see Sec.~\ref{sec:keyboard_shortcuts}.
\item Gradient flow and non-arrested Newton flow.
\item Snapshot saver using \texttt{Ctrl+p} to save the current
  window's visualization of the current Skyrmion to a \texttt{.png}
  file without using third-party software.
\item File-load functionality that can load and distinguish the text
  format and the binary format automatically, for quickly continuing a
  Skyrmion computation from saved data.
\end{itemize}


\subsection{GPU execution model and \texttt{Numba CUDA}}

The computational core of \texttt{skyrmion\_solver} is written in \texttt{Python} using \texttt{Numba CUDA}.
The field theory is therefore expressed directly as compiled CUDA kernels, typically through the decorators \verb|@cuda.jit(device=True)| for device helpers and \verb|@cuda.jit| for launchable kernels \cite{Cautaerts_2026}.
This keeps the implementation close in spirit to \texttt{cuSkyrmion}: the local PDE terms are written explicitly by hand, rather than hidden behind a higher-level tensor framework.

The basic parallel decomposition is one lattice site per CUDA thread on a structured three-dimensional finite-difference grid.
The execution model is single instruction, multiple threads (SIMT): the same kernel is applied pointwise across the lattice, with each thread evaluating the local update of the fields at one site.
CUDA launches are organized as three-dimensional grids of thread blocks, each block covering a local tile of the simulation domain.
Global lattice coordinates are recovered from \verb|blockIdx|, \verb|threadIdx|, and \verb|blockDim|, or equivalently through the convenient \texttt{Numba} idiom
\begin{verbatim}
x, y, z = cuda.grid(3)
\end{verbatim}
after which threads lying outside the active domain are discarded by explicit bounds checks.

The dominant numerical pattern is the stencil.
Each thread reads a fixed local neighbourhood, evaluates high-order finite-difference approximations to first and second derivatives, and assembles the local PDE terms entering the energy density, baryon density, gradient flow, or observable under consideration.
Most kernels operate directly out of global memory, since the full three-dimensional field volumes are too large to reside in on-chip storage.
Shared memory is used selectively for block-local cooperation, primarily in reduction kernels rather than as the main storage mechanism for the field variables.
In particular, sums, maxima, and minima are first reduced within a thread block in shared memory and then written to global memory as one partial result per block.
Synchronization within the block is handled with \verb|cuda.syncthreads()| between reduction stages.

Time evolution and relaxation follow the same explicit kernel-launch structure.
Fourth-order Runge--Kutta is implemented as a sequence of CUDA launches over the full lattice, while arrested Newton flow uses the same infrastructure together with an energy-based arrest criterion to reset the fictitious velocity when the trial step overshoots.
Kernel launches are asynchronous with respect to the \texttt{Python} host by default, so synchronization is introduced only when values are needed immediately for diagnostics, reductions, or rendering.
In this way, \texttt{Numba CUDA} exposes the CUDA execution model directly while still allowing the surrounding framework to be written in \texttt{Python}.


\subsection{CUDA--OpenGL interoperability and rendering}

A second major component inherited from \texttt{cuSkyrmion} is the close integration of computation and visualization.
In \texttt{skyrmion\_solver}, low-level CUDA driver interaction is handled through \texttt{cuda-python}, in particular for CUDA--OpenGL interoperability.
The OpenGL side of the rendering pipeline is implemented through \texttt{PyOpenGL}, while \texttt{glfw} provides context creation, window management, and interactive input handling.

The rendering pipeline keeps the main visualization loop GPU resident.
OpenGL pixel buffer objects are registered with CUDA and mapped into device address space for each frame.
CUDA kernels write RGBA volume data directly into these mapped buffers without staging through host memory.
The resulting volume is uploaded to a three-dimensional texture and displayed by real-time CUDA--OpenGL volume ray-tracing \cite{Storti_2015}.
Baryon density is typically used to control opacity, while the normalized pion vector or a related field-derived quantity controls the colour.
This preserves one of the most useful features of \texttt{cuSkyrmion}: the ability to inspect the development of symmetry, clustering, and metastability in real time while the relaxation is running.


\subsection{Supported models and modular theory interface}

The present implementation of \texttt{skyrmion\_solver} is centred on three-dimensional Skyrme-type effective field theories for baryons and nuclei.
The broader motivation comes from the large-$N_{\rm c}$ perspective of QCD, where mesonic effective theories acquire a natural solitonic interpretation and baryons emerge as topological excitations rather than as fundamental fields \cite{Hooft_1974,Witten_1979}.
In the original proposal of Skyrme, the basic degree of freedom is an $\SU(2)$-valued chiral field, and the baryon number is identified with its topological degree \cite{Skyrme_1961,Manton_2022}.
That basic picture remains the common starting point of all built-in theories in the package.

At the numerical level, each theory is implemented through the same software interface.
A theory module specifies its field content, parameter system, local energy density, associated Euler--Lagrange equations, initialization routines, observables, and any theory-specific rendering helpers.
The theory registry then allows the model to be loaded at runtime and coupled to the common \texttt{Simulation} driver.
This is precisely the point of the framework design: the execution model remains fixed, while the physics is supplied by the selected theory module.

The currently supported theories are listed in the following subsubsections.
They all share the same basic computational workflow, but differ substantially in physical interpretation.
Some modify the pion sector itself, as in the lightly bound
\cite{Harland:2013rxa,Gillard:2015eia} and Berger-type
models \cite{Speight_2021}.
Others enlarge the field content by introducing additional mesonic or
electromagnetic degrees of freedom, as in the $\rho$-meson
\cite{Adkins_1986,Leask_Naya_2025}, $\omega$-meson
\cite{Nappi_1984,Sutcliffe_2009,Gudnason_2020,Leask_Harland_2024}, and
Coulomb-coupled variants \cite{Gudnason_2025}.
This makes the package useful not only for isolated multi-Skyrmions,
but also for crystal phases of dense matter, low-binding-energy
variants, and backreacted coupled systems relevant to nuclear
structure and neutron-star applications.
Except for the Berger-Skyrme variant, all the Skyrme-model
  variants have been studied previously in the literature and here we
  simply review them for the implementation 
  in the CUDA GPU code, \texttt{skyrmion\_solver}.
  The geometric formulation of the Berger-Skyrme model
  \cite{Speight_2021} has not been previously published in the
  literature and it is physically and mathematically different from a
  similar model of Ward
  \cite{Ward_2004,SilvaLobo_2011,Naya_2021}.
  Indeed, the model of Ward interpolates between the Skyrme model and
  the Faddeev-Skyrme model also by squashing the target 3-sphere, but
  reaches the Faddeev-Skyrme model in a degenerate limit where the
  target space degenerates into the 2-sphere, whereas the
  Berger-Skyrme model does not reduce to the latter although the
  target space is also squashed.
  The two models hence differ both in symmetries and in energy
  functionals.


\subsubsection{The standard massive nuclear Skyrme model}

The standard massive nuclear Skyrme model is the basis of the \texttt{cuSkyrmion} code and is described in Sec.~\ref{sec: Skyrme model}, so we will not repeat the description in its entirety here for the \texttt{skyrmion\_solver} code.
The Lagrangian in physical units is given in Eq.~\eqref{eq: Skyrme Lagrangian} and in dimensionless units in Eq.~\eqref{eq:Lag_dimensionless}.
The static energy density is given in Eq.~\eqref{eq: Skyrme model - Static energy} in dimensionless units and finally the equation of motion in Eq.~\eqref{eq:skyrme-static-eom}.
The energy and length units for the dimensionless energy, Lagrangian and equations of motion are defined in Eq.~\eqref{eq:energy_length_units}.

\begin{figure*}[t]
    \centering
    \begin{subfigure}[b]{0.45\textwidth}
        \includegraphics[width=\textwidth]{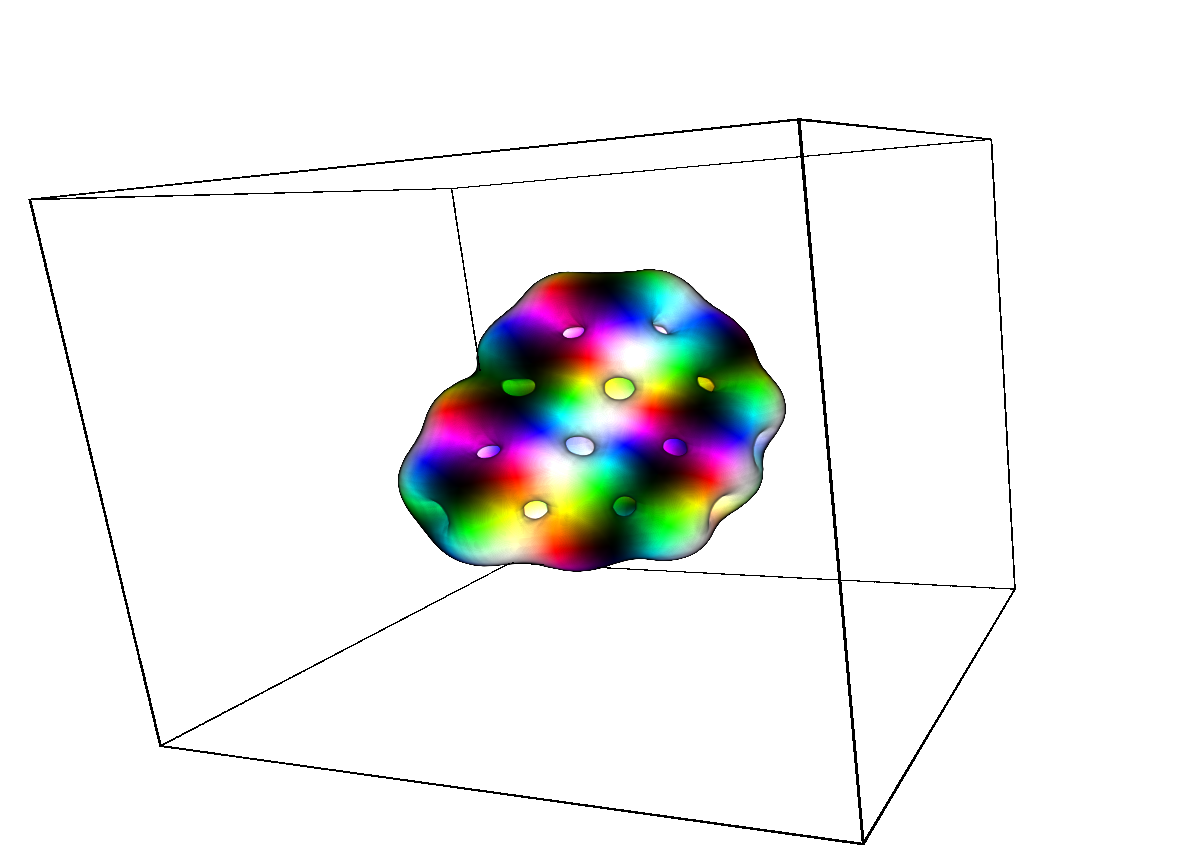}
        \caption{Standard massive nuclear Skyrme model}
        \label{fig: nuclear}
    \end{subfigure}
    ~
    \begin{subfigure}[b]{0.45\textwidth}
        \includegraphics[width=\textwidth]{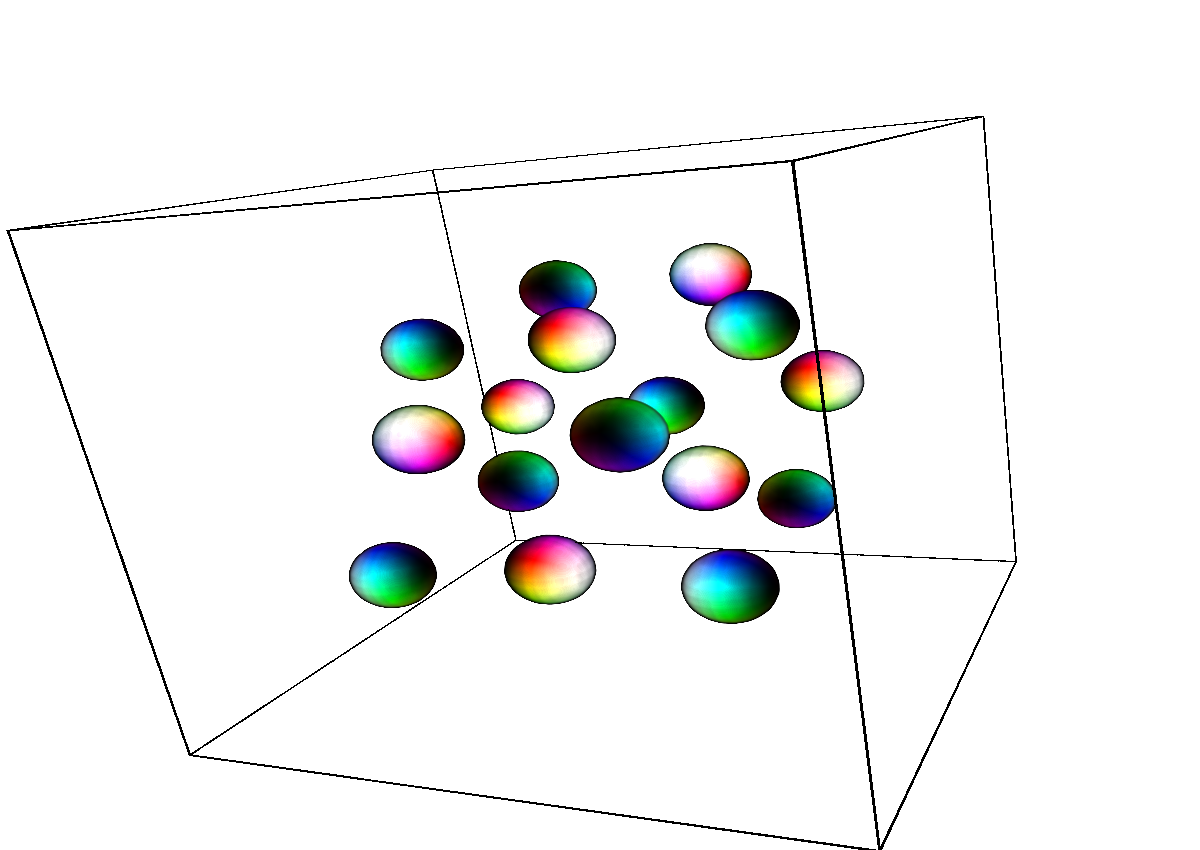}
        \caption{Lightly bound Skyrme model ($\alpha=0.95$)}
        \label{fig: lightly}
    \end{subfigure}
    \caption{$B=16$ massive Skyrmions, with dimensionless mass $m=1$, in the (a) standard massive nuclear and (b) lightly bound Skyrme models, obtained in \texttt{skyrmion\_solver} using the initialization \texttt{sim.initialize(\{"mode":"smorgasbord", "baryon\_number":16, "seed":3\})}.}
    \label{fig: B=16 skyrmions}
\end{figure*}

This is the reference theory against which the remaining variants and extensions are compared.
An example $B=16$ Skyrmion obtained using \texttt{skyrmion\_solver} is shown in Fig.~\ref{fig: B=16 skyrmions}.

\subsubsection{The lightly bound Skyrme model}

Although the standard massive Skyrme model captures many qualitative features of nuclear physics, it notoriously overestimates classical binding energies.
This motivates deformations of the pion theory that preserve the topological soliton picture while modifying the balance of attractive and repulsive contributions in the static energy.
One particularly successful proposal is the lightly bound Skyrme model \cite{Harland:2013rxa,Gillard:2015eia,Gillard_2017}.
The field remains $U : \mathbb{R}^{1,3} \to \SU(2)$, but the potential sector is modified in such a way as to favour more weakly bound, particle-like multi-Skyrmion configurations.

In physical units, the Lagrangian is
\begin{align}
\label{eq:lb-physical-lagrangian}
    \mathcal{L}_{\textup{Lig}}[U] = \, & (1-\alpha) \left( \frac{F_\pi^2}{16\hbar}\Tr(L_\mu L^\mu) - \frac{F_\pi^2m_\pi^2}{8\hbar^3}\Tr(\Id_2-U)
    \right) + \frac{\hbar}{32g^2}\Tr\!\left([L_\mu,L_\nu][L^\mu,L^\nu]\right) \nonumber \\
    \, & - \frac{\alpha F_\pi^4 g^2}{512\hbar^3(1-\alpha)^2}\Tr(\Id_2-U)^4.
\end{align}
The first three terms are the standard massive Skyrme terms, except that the sigma-model and pion-mass terms are scaled by $(1-\alpha)$.
The final term is the holomorphic quartic pion potential, known as the lightly bound potential \cite{Gudnason:2016mms}.
It is this additional potential that drives the model into a low-binding energy regime.
In particular, for suitable values of $\alpha$ the model favours clustered multi-Skyrmions, and this leads to a much better account of classical nuclear binding energies than in the standard theory.

The corresponding static energy is
\begin{align}
\label{eq:lb-physical-energy}
    E_{\textup{Lig}}[U] = \, & \int_{\mathbb{R}^3}\d^3x
    \Bigg\{ (1-\alpha) \left( -\frac{F_\pi^2}{16\hbar}\Tr(L_iL_i) + \frac{F_\pi^2m_\pi^2}{8\hbar^3}\Tr(\Id_2-U) \right) \nonumber\\
    & - \frac{\hbar}{32g^2}\Tr\!\left([L_i,L_j][L_i,L_j]\right) + \frac{\alpha F_\pi^4 g^2}{512\hbar^3(1-\alpha)^2}\Tr(\Id_2-U)^4 \Bigg\}.
\end{align}
It is natural in this case to use the rescaled units
\begin{equation}
    \tilde E = \frac{F_\pi}{4g}\sqrt{1-\alpha},
    \quad
    \tilde L = \frac{2\hbar}{gF_\pi}\sqrt{1-\alpha},
\end{equation}
for which the dimensionless pion mass becomes
\begin{equation}
    m = \frac{2m_\pi\sqrt{1-\alpha}}{gF_\pi}.
\end{equation}
Then the static energy takes the dimensionless form
\begin{align}
\label{eq:lb-dimensionless-energy}
    E_{\textup{Lig}}[U] = \, & \int_{\mathbb{R}^3}\d^3x \Bigg\{ (1-\alpha) \left[-\frac{1}{2}\Tr(L_iL_i) + m^2\Tr(\Id_2-U) \right] - \frac{1}{16}\Tr\!\left([L_i,L_j][L_i,L_j]\right) \nonumber \\
    \, & + \alpha \left( \frac{1}{2}\Tr(\Id_2-U) \right)^4 \Bigg\}.
\end{align}
The static Euler--Lagrange equation is again obtained by varying through $U \mapsto U e^\varepsilon$.
It may be written as
\begin{equation}
\label{eq:lb-static-eom}
    \partial_i \left( (1-\alpha)L_i + \frac{1}{4}[L_j,[L_i,L_j]] \right) - \frac{(1-\alpha)m^2}{2}(U-U^\dagger) - \alpha \, \left[ \Tr(\Id_2-U)\right]^3(U-U^\dagger) = 0.
\end{equation}
The structure is therefore very close to that of the standard model, but with a substantially altered potential sector.
It is precisely this modification that gives rise to the improved low-binding phenomenology.
A $B=16$ lightly bound Skyrmion is shown alongside a massive standard Skyrmion in Fig.~\ref{fig: B=16 skyrmions}.


\subsubsection{The Berger-Skyrme model}

A different deformation of the standard Skyrme theory is obtained by modifying the geometry of the target space itself.
Rather than equipping the target $\SU(2)\cong S^3$ with the round bi-invariant metric, one introduces a one-parameter family of left-invariant metrics which squash the Hopf fibres relative to the base $S^2$.
This yields the Berger-Skyrme model.
It is closely related in spirit to the squashed-sphere models studied by Ward and by Silva Lobo and Ward, which interpolate between the ordinary Skyrme model and the Skyrme--Faddeev system \cite{Ward_2004,SilvaLobo_2011,Naya_2021}.
The present model differs from those constructions in that we remain
throughout with an $\SU(2)$-valued Skyrme field and do not take the
degenerate Hopf limit.
Indeed although the kinetic term in the Berger-Skyrme model
  reduces identically to that of the Faddeev-Skyrme model in the $\alpha\to0$
  limit, the Skyrme term reduces to the exact complement of that
  of the Faddeev-Skyrme model.
Instead, we regard the target as a squashed $S^3$ and study the resulting anisotropic Skyrme theory in its own right.

Let the Skyrme field be the map $U: \mathbb{R}^3 \to \SU(2)$.
Write $T_a=-i\tau_a$ for the standard basis of $\su(2)$, where $\tau_a$ are the Pauli matrices.
Let $\{\theta_a\}$ denote the corresponding left-invariant vector fields on $\SU(2)$, and let $\{\sigma_a\}$ be the dual left-invariant one-forms, so that $\sigma_a(\theta_b)=\delta_{ab}$.
A general left-invariant metric on $\SU(2)$ may be diagonalized in a Milnor frame.
The Berger family is obtained by taking
\begin{equation}
    h_\alpha = \sigma_1^2+\sigma_2^2+\alpha^2\sigma_3^2,
\label{eq: Berger Skyrmions - Berger metric}
\end{equation}
so that distances along the Hopf $S^1$-fibres are scaled by the factor $\alpha$.
Equivalently,
\begin{equation}
    h_\alpha(T^a,T^b)=\delta^{ab}+(\alpha^2-1)\delta^{3a}\delta^{3b}.
\end{equation}
When $\alpha=1$ the target is the round three-sphere and one recovers the standard Skyrme model.
For $\alpha\neq1$, one internal direction is distinguished and the full $\SO(4)$ symmetry of the round target is reduced to the subgroup preserving the Hopf-fibre structure, naturally identified with $\U(2)$.

From the geometric point of view, the quadratic and quartic parts of the static energy are most naturally written directly in terms of pullbacks of the left-invariant forms \cite{Speight_2021}:
\begin{subequations}
    \begin{align}
        E_2[U] & = c_2 \int_{\mathbb{R}^3} \d^3x \left\{ |U^* \sigma_1|^2 + |U^* \sigma_2|^2 + \alpha^2 |U^* \sigma_3|^2 \right\}, \\
        E_4[U] & = c_4 \int_{\mathbb{R}^3} \d^3x \left\{ |U^* (\sigma_2 \wedge \sigma_3)|^2 + |U^* (\sigma_3 \wedge \sigma_1)|^2 + \alpha^2 |U^* (\sigma_1 \wedge \sigma_2)|^2 \right\}.
    \end{align}
\end{subequations}
These formulas make the target-space anisotropy completely explicit.
The Hopf-fibre direction $\sigma_3$ is weighted differently already in the Dirichlet term, and the same deformation propagates to the quartic Skyrme term through the norms of the pulled-back area forms.
This is precisely the natural Skyrme energy associated with the Berger metric on the target.

To connect this formulation with the usual field-theoretic notation, write the field in the $\sigma$-model notation as
\begin{equation}
    U = \varphi^0 \Id_2 + i\varphi^a\tau^a,
\end{equation}
where $\varphi=(\varphi^0,\varphi^1,\varphi^2,\varphi^3)\in S^3\subset\mathbb{R}^4$.
The pullback of the Maurer--Cartan form is
\begin{equation}
    L_i = U^\dagger \partial_i U = L_i^a T_a,
\end{equation}
and the curvature two-form is
\begin{equation}
    \Omega_{ij}=[L_i,L_j]=2\Omega_{ij}^aT_a.
\end{equation}
In these variables one finds
\begin{align}
    h_\alpha(L_i,L_j) &= L_i^aL_j^a + (\alpha^2-1)L_i^3L_j^3,\\
    h_\alpha(\Omega_{ij},\Omega_{kl}) &= 4\Omega_{ij}^a\Omega_{kl}^a + 4(\alpha^2-1)\Omega_{ij}^3\Omega_{kl}^3.
\end{align}
Thus the Berger deformation may be viewed as the standard round-target theory supplemented by terms involving only the third internal component.
This is the form used in the field-theoretic Lagrangian.

A particularly useful way to express these distinguished third components is through a fixed symplectic form on $\mathbb{R}^4$.
A direct calculation gives
\begin{align}
    L_i^3
    &=
    \partial_i \varphi^1 \varphi^2 - \partial_i \varphi^2 \varphi^1 + \partial_i \varphi^0 \varphi^3 - \partial_i \varphi^3 \varphi^0 \nonumber\\
    &= \omega_{\alpha\beta}\partial_i\varphi^\alpha \varphi^\beta,
\end{align}
and similarly
\begin{align}
    \Omega_{ij}^3
    &=
    \partial_i \varphi^1 \partial_j \varphi^2 - \partial_i \varphi^2 \partial_j \varphi^1 + \partial_i \varphi^0 \partial_j \varphi^3 - \partial_i \varphi^3 \partial_j \varphi^0 \nonumber\\
    &= \omega_{\alpha\beta}\partial_i\varphi^\alpha \partial_j\varphi^\beta,
\end{align}
where $\alpha,\beta\in\{0,1,2,3\}$, and the symplectic matrix $\omega$ is
\begin{equation}
    (\omega_{\alpha\beta})=
    \begin{pmatrix}
    0 & 0 & 0 & +1 \\
    0 & 0 & +1 & 0 \\
    0 & -1 & 0 & 0 \\
    -1 & 0 & 0 & 0
    \end{pmatrix}.
\end{equation}
Hence the anisotropic pieces may be written compactly as
\begin{equation}
    L_i^3L_j^3
    =
    \omega_{\alpha\beta}\omega_{\gamma\delta}
    \partial_i\varphi^\alpha \varphi^\beta
    \partial_j\varphi^\gamma \varphi^\delta,
\end{equation}
and
\begin{equation}
    \Omega_{ij}^3\Omega_{kl}^3
    =
    \omega_{\alpha\beta}\omega_{\gamma\delta}
    \partial_i\varphi^\alpha \partial_j\varphi^\beta
    \partial_k\varphi^\gamma \partial_l\varphi^\delta.
\end{equation}
The appearance of this fixed symplectic form is characteristic of the Berger deformation and provides a convenient algebraic encoding of the Hopf-fibre anisotropy in sigma-model variables.

The skew matrix $\omega_{\alpha\beta}$ is naturally interpreted as the coefficient matrix of a constant symplectic $2$-form on the ambient space $\mathbb{R}^4$ containing $S^3$. In the coordinates $\varphi=(\varphi^0,\varphi^1,\varphi^2,\varphi^3)$, one may write $\omega = \d\varphi^0 \wedge \d\varphi^3 + \d\varphi^1 \wedge \d\varphi^2$, and then the distinguished Berger quantity $L_i^3$ is just the contraction $\omega(\varphi,\partial_i\varphi)$. Thus the preferred $\sigma_3$ direction is not arbitrary, but is induced by the ambient symplectic geometry. Since $S^3$ is odd-dimensional it is not itself symplectic, but the restriction of $\iota_r\omega$, where $r$ is the radial vector field on $\mathbb{R}^4$, defines the standard contact form on $S^3$ whose Reeb direction is precisely the Hopf fibre. In this sense, the Berger deformation is geometrically tied to the Hopf fibration through the ambient symplectic structure.

In physical units the resulting field-theoretic Lagrangian may be written as
\begin{align}
\label{eq:berger-physical-lagrangian}
    \mathcal{L}_{\textup{Ber}}[U] = \, &\frac{F_\pi^2}{16\hbar}\Tr(L_\mu L^\mu) + \frac{\hbar}{32g^2}\Tr\!\left([L_\mu,L_\nu][L^\mu,L^\nu]\right) - \frac{F_\pi^2m_\pi^2}{8\hbar^3}\Tr(\Id_2-U) \nonumber \\
    \, &- (\alpha^2-1) \left( \frac{F_\pi^2}{8\hbar}L_\mu^3L^{3\mu} + \frac{\hbar}{4g^2}\Omega_{\mu\nu}^3\Omega^{3\mu\nu}\right),
\end{align}
where the anisotropic corrections are precisely those induced by the Berger metric.
The static energy functional is therefore
\begin{align}
\label{eq:berger-physical-energy}
    E_{\textup{Ber}}[U]
    =
    \int_{\mathbb{R}^3}\d^3x
    \Bigg\{
        &-\frac{F_\pi^2}{16\hbar}\Tr(L_iL_i)
        -
        \frac{\hbar}{32g^2}\Tr\!\left([L_i,L_j][L_i,L_j]\right)
        +
        \frac{F_\pi^2m_\pi^2}{8\hbar^3}\Tr(\Id_2-U)
        \nonumber\\
        &
        +
        (\alpha^2-1)
        \left(
            \frac{F_\pi^2}{8\hbar}L_i^3L_i^3
            +
            \frac{\hbar}{4g^2}\Omega_{ij}^3\Omega_{ij}^3
        \right)
    \Bigg\}.
\end{align}

\begin{figure*}[t]
    \centering
    \begin{subfigure}[b]{0.45\textwidth}
        \includegraphics[width=\textwidth]{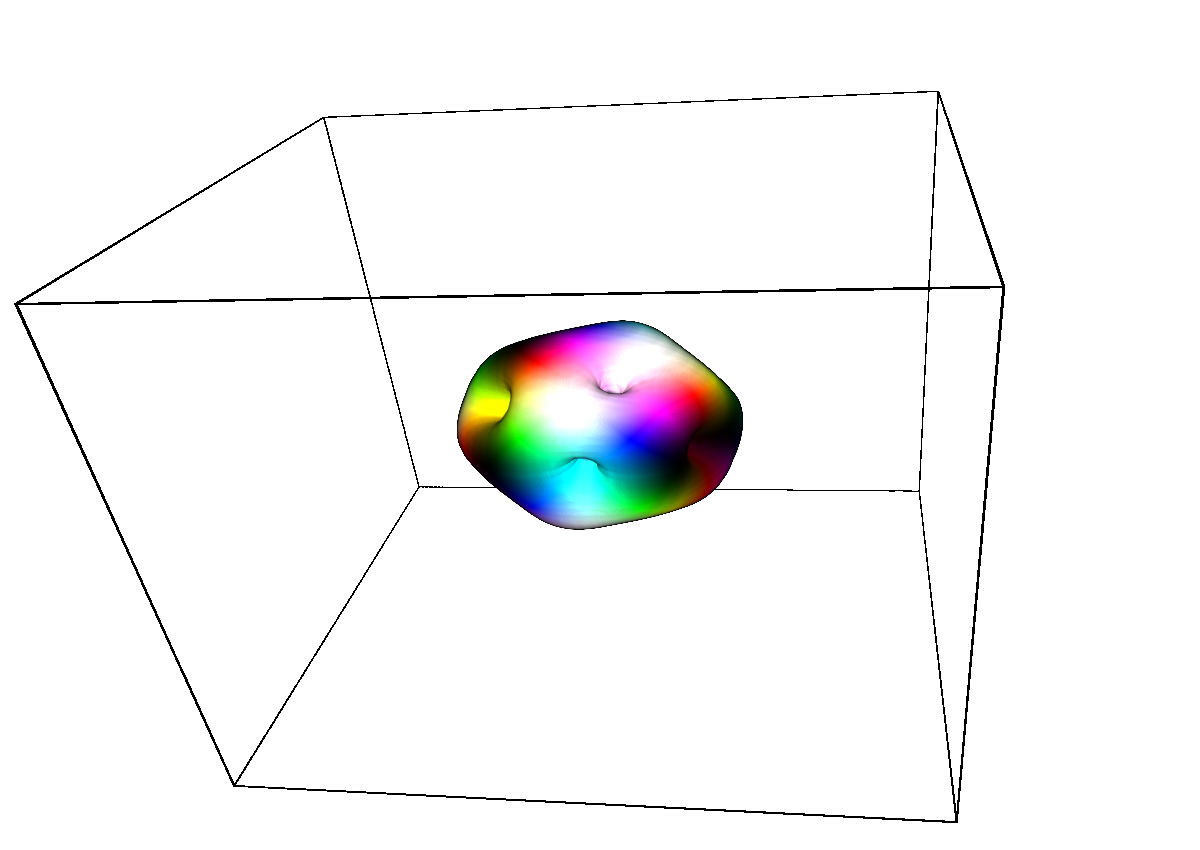}
        \caption{Standard ($\alpha=1$)}
        \label{fig: Berger 1}
    \end{subfigure}
    ~
    \begin{subfigure}[b]{0.45\textwidth}
        \includegraphics[width=\textwidth]{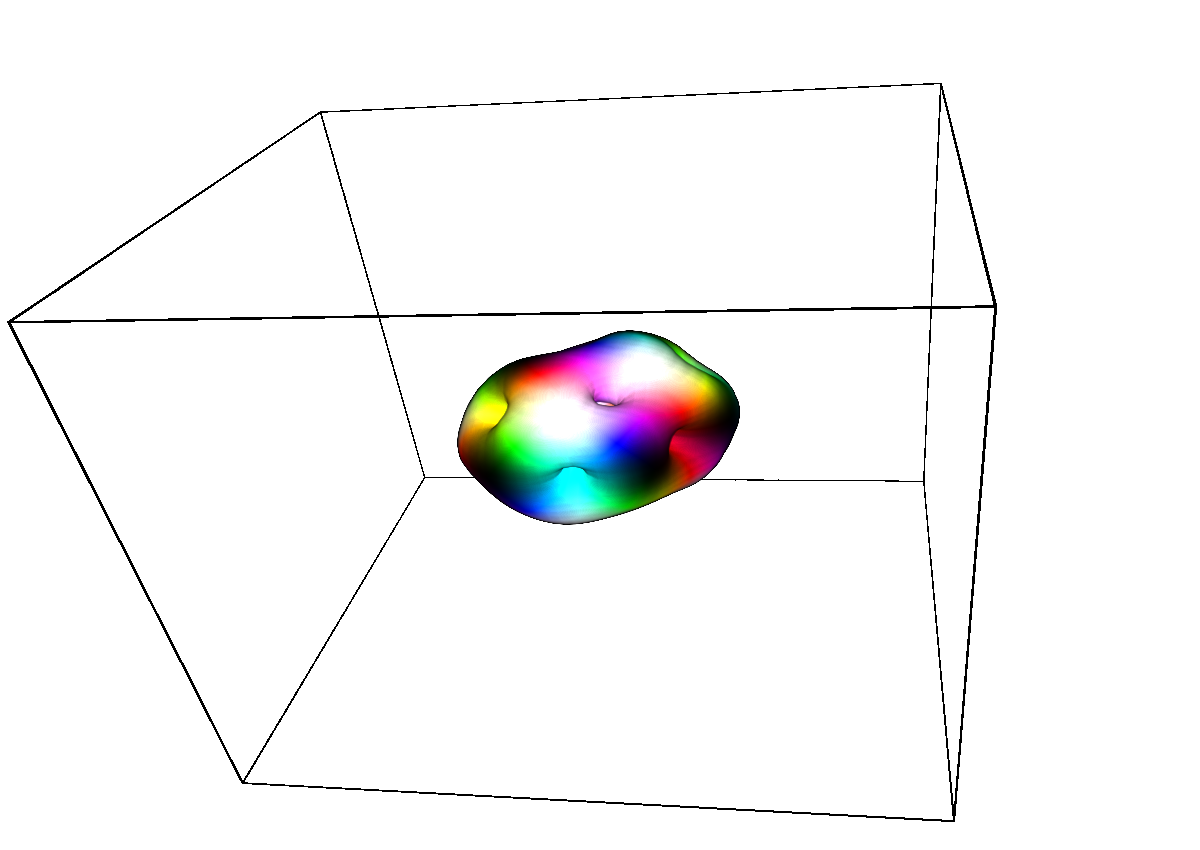}
        \caption{Intermediate ($\alpha=0.5$)}
        \label{fig: Berger 0.5}
    \end{subfigure}
    ~
    \begin{subfigure}[b]{0.45\textwidth}
        \includegraphics[width=\textwidth]{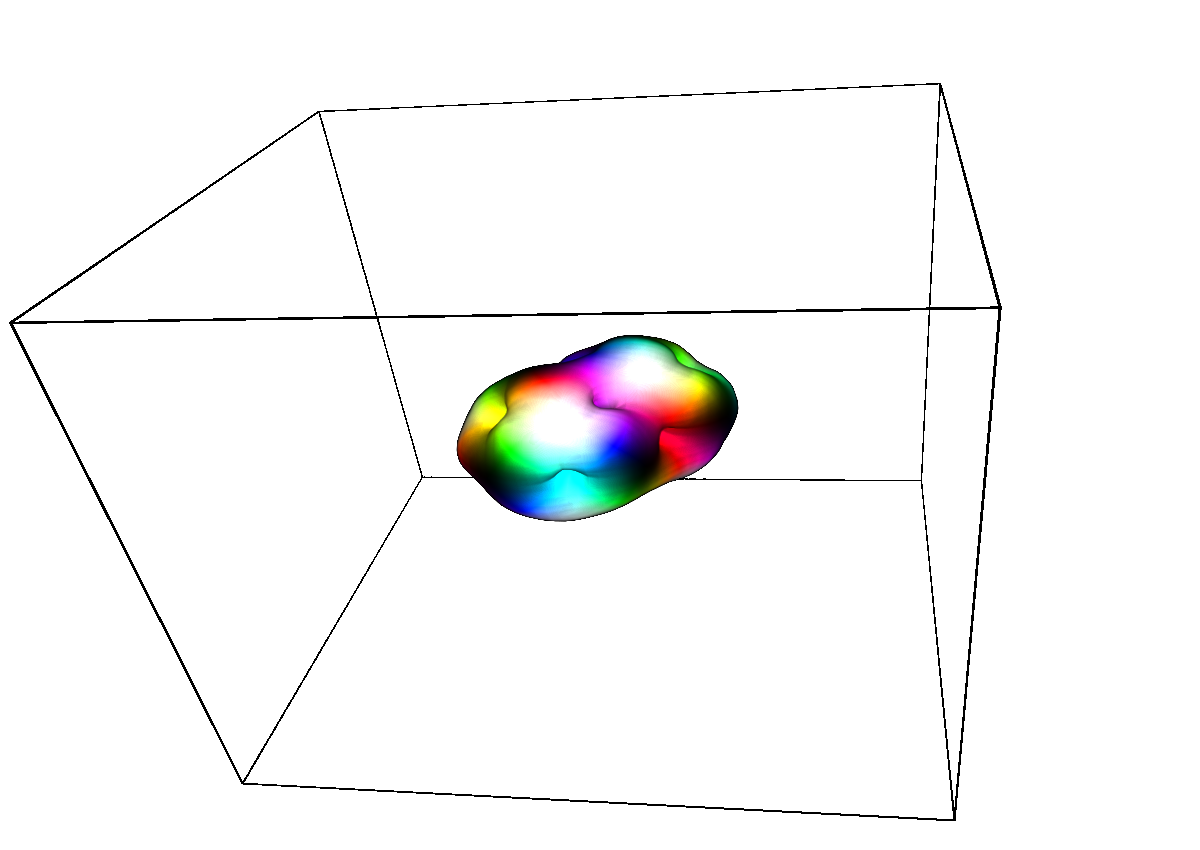}
        \caption{BPS ($\alpha=0$)}
        \label{fig: Berger 0}
    \end{subfigure}
    \caption{$B=7$ Skyrmion in the Berger-Skyrme model, with dimensionless pion mass $m=1$, for (a) $\alpha=1$, (b) $\alpha=0.5$, and (c) $\alpha=0$.}
    \label{fig: B=7 Berger}
\end{figure*}

Using the same Skyrme units as in the standard model
$\tilde E = \frac{F_\pi}{4g}$,
$\tilde L = \frac{2\hbar}{gF_\pi}$,
and $m = \frac{2m_\pi}{gF_\pi}$,
one obtains the dimensionless static energy
\begin{align}
\label{eq:berger-dimensionless-energy}
    E_{\textup{Ber}}[U]
    =
    \int_{\mathbb{R}^3}\d^3x
    \Bigg\{
        &-\frac{1}{2}\Tr(L_iL_i)
        -
        \frac{1}{16}\Tr\!\left([L_i,L_j][L_i,L_j]\right)
        +
        m^2\Tr(\Id_2-U)
        \nonumber\\
        &
        +
        (\alpha^2-1)
        \left(
            L_i^3L_i^3
            +
            \frac{1}{2}\Omega_{ij}^3\Omega_{ij}^3
        \right)
    \Bigg\}.
\end{align}
This is the form implemented in the software.

The corresponding Euler--Lagrange equation is obtained by varying the constrained field $U\in\SU(2)$.
It may be written schematically as
\begin{equation}
\label{eq:berger-static-eom}
    \partial_i
    \left(
        L_i + \frac{1}{4}[L_j,[L_i,L_j]]
        +(\alpha^2-1)L_i^3 T_3
        + \mathcal{J}_i^{\textup{Ber}}
    \right)
    - \frac{m^2}{2}(U-U^\dagger)
    = 0,
\end{equation}
where $\mathcal{J}_i^{\textup{Ber}}$ denotes the contribution from variation of the $\Omega_{ij}^3\Omega_{ij}^3$ term.
The essential point is that the deformation leaves the topological degree unchanged, but modifies the target-space geometry and hence both the local field equations and the detailed structure of the static solutions.
The effect of the factor $\alpha$ on the $B=7$ Skyrmion is shown in
Fig.~\ref{fig: B=7 Berger} and we have explicitly checked
  that the $\alpha=1$ case is identical to the solution found using
  the standard (nuclear) Skyrme model code.

It is worth emphasizing once more how this differs from the Ward--Silva Lobo family.
In those models the squashing parameter interpolates all the way to the degenerate $\CP^1$ limit, where one recovers the Skyrme--Faddeev system and Hopf solitons \cite{Ward_2004,SilvaLobo_2011}.
Here, by contrast, the target remains $\SU(2)$ and the model is used as an anisotropic Skyrme theory.
The Berger deformation is therefore best viewed as a symmetry-breaking modification of the standard Skyrme model induced by a squashed $S^3$ target metric, not as a full reduction to the Hopfion sector.


\subsubsection{The \texorpdfstring{$\rho$}{rho}-meson extension}

It is natural to ask whether the quartic Skyrme term should really be regarded as a fundamental ingredient of the theory, or whether it should instead be understood as an effective remnant of heavier mesonic degrees of freedom.
From the large-$N_c$ point of view, the low-energy effective theory of QCD is mesonic, not merely pionic, and vector mesons should therefore appear on the same conceptual footing as the pions \cite{Hooft_1974,Witten_1979}.
This idea underlies a number of vector-meson extensions of the Skyrme model, including hidden local symmetry constructions and holographic reductions \cite{Igarashi_1985,Meissner_1986,Forkel_1991,Park_2004,Yong-Liang_2013,Sutcliffe_2010}.
The $\rho$-meson is especially natural in this setting because it is an isovector field and hence couples directly to the pion current sector.

Coupling the Skyrme model in a chirally invariant way to $\rho$-mesons was first proposed by Adkins \cite{Adkins_1986}.
An alternative model was proposed by Meissner where the Skyrme term was replaced by the sextic term \cite{Meissner_1987}.
In both of these models, the $\rho$-meson is treated as a constrained $2\times 2$ four vector.
However, the $\rho$-meson in its standard form is a massive non-Abelian field.
In the present model we have the usual Skyrme field $U \in \SU(2)$ and the $\rho$-meson $R_\mu \in \su(2)$ with curvature
\begin{equation}
    R_{\mu\nu} = \partial_\mu R_\nu - \partial_\nu R_\mu,
\end{equation}
where $R_\mu=i \rho^a_\mu \tau^a$ with $\rho^a_\mu \in \mathbb{R}$ and $\tau^a$ are the Pauli spin matrices.
A convenient physical Lagrangian was proposed in Ref.~\cite{Leask_Naya_2025}, and is given by
\begin{align}
\label{eq:rho-physical-lagrangian}
    \mathcal{L}_{\rho}[U,R_\mu] = \, & \frac{F_\pi^2}{16\hbar}\Tr(L_\mu L^\mu) + \frac{\hbar}{32g^2}\Tr\!\left([L_\mu,L_\nu][L^\mu,L^\nu]\right) - \frac{F_\pi^2m_\pi^2}{8\hbar^3}\Tr(\Id_2-U) \nonumber\\
    \, & - \frac{m_\rho^2}{4\hbar^3}\Tr(R_\mu^\dagger R^\mu) - \frac{1}{8\hbar}\Tr(R_{\mu\nu}^\dagger R^{\mu\nu}) + \frac{1}{2}\alpha \, \eta^{\mu\beta}\eta^{\nu\gamma}\Tr\!\left(R_{\mu\nu}[L_\beta,L_\gamma]\right).
\end{align}
This particular interaction is motivated both by holographic-like reductions of Yang--Mills theory and by the fact that it generates the physical $\rho\pi\pi$ vertex in the low-energy limit.
In the simplified form used here, one keeps a minimal $\rho$-meson coupling rather than the full tower of interaction terms appearing in a more systematic reduction \cite{Sutcliffe_2018}.

\begin{figure*}[t]
    \centering
    \begin{subfigure}[b]{0.45\textwidth}
        \includegraphics[width=\textwidth]{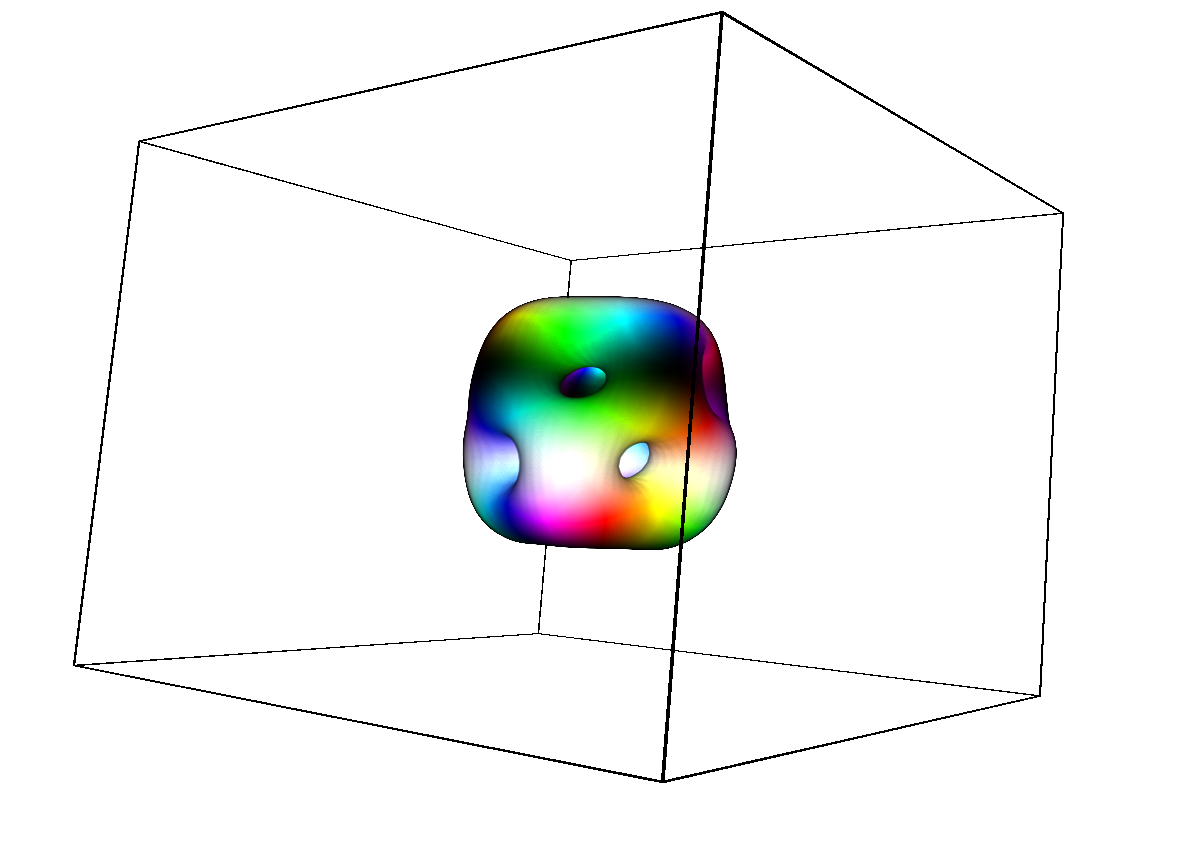}
        \caption{Baryon density}
        \label{fig: rho - runge}
    \end{subfigure}
    ~
    \begin{subfigure}[b]{0.45\textwidth}
        \includegraphics[width=\textwidth]{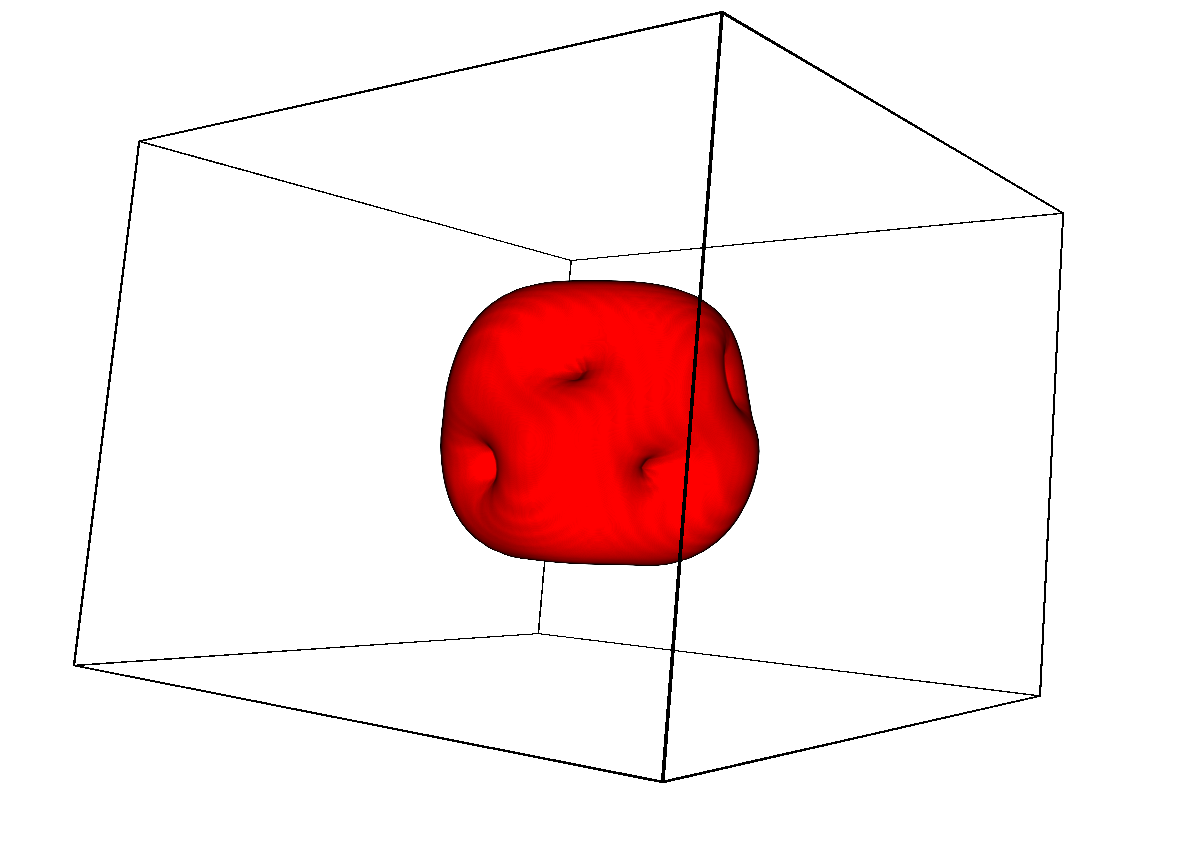}
        \caption{$\rho$-meson density, $|\rho_\mu^a|^2$}
        \label{fig: rho - rho}
    \end{subfigure}
    \caption{$B=6$ Skyrmion in the $\rho$-Skyrme model, showing (a) the baryon density and (b) the $\rho$-meson density.}
    \label{fig: rho}
\end{figure*}

For static fields the energy is
\begin{align}
\label{eq:rho-physical-energy}
    E_{\rho}[U,R_i] = \, & \int_{\mathbb{R}^3}\d^3x \Bigg\{-\frac{F_\pi^2}{16\hbar}\Tr(L_iL_i) - \frac{\hbar}{32g^2}\Tr\!\left([L_i,L_j][L_i,L_j]\right) + \frac{F_\pi^2m_\pi^2}{8\hbar^3}\Tr(\Id_2-U) \nonumber\\
    & + \frac{m_\rho^2}{4\hbar^3}\Tr(R_i^\dagger R_i) + \frac{1}{8\hbar}\Tr(R_{ij}^\dagger R_{ij}) - \frac{1}{2}\alpha \Tr\!\left(R_{ij}[L_i,L_j]\right)\Bigg\}.
\end{align}
Passing to Skyrme units $\tilde E = \frac{F_\pi}{4g}$ and $\tilde L = \frac{2\hbar}{gF_\pi}$, and writing
\begin{equation}
    M_\pi = \frac{2m_\pi}{gF_\pi},
    \quad
    M_\rho = \frac{2m_\rho}{gF_\pi},
    \quad
    c_\alpha = \frac{\alpha g F_\pi}{4},
\end{equation}
one obtains the dimensionless static energy
\begin{align}
\label{eq:rho-dimensionless-energy}
    E_{\rho}[U,R_i] = \, & \int_{\mathbb{R}^3}\d^3x \Bigg\{ -\frac{1}{2}\Tr(L_iL_i) - \frac{1}{16}\Tr\!\left([L_i,L_j][L_i,L_j]\right) + M_\pi^2\Tr(\Id_2-U) \nonumber\\
    \, & + M_\rho^2\Tr(R_i^\dagger R_i) + \frac{1}{2}\Tr(R_{ij}^\dagger R_{ij}) - 4 c_\alpha \Tr\!\left(R_{ij}[L_i,L_j]\right) \Bigg\}.
\end{align}
The static Euler--Lagrange equations are now a coupled system,
\begin{equation}
\label{eq:rho-static-eom-U}
    \partial_i \left(L_i + \frac{1}{4}[L_j,[L_i,L_j]] - 8c_\alpha [R_{ij},L_j] \right) - \frac{M_\pi^2}{2}(U-U^\dagger) = 0,
\end{equation}
and
\begin{equation}
\label{eq:rho-static-eom-R}
    \partial_j R_{ji} - M_\rho^2 R_i + 4c_\alpha \partial_j([L_j,L_i]) = 0.
\end{equation}
The first equation is a Skyrme-type equation with explicit $\rho$-meson backreaction, while the second is a massive Proca-type equation driven by the pion current sector.
A $B=6$ soliton in the $\rho$-Skyrme model is shown in Fig.~\ref{fig: rho}.

This model also clarifies the status of the quartic Skyrme term.
If one neglects derivatives of $R_i$ and formally takes the infinite-mass limit $M_\rho \to \infty$, then Eq.~\eqref{eq:rho-static-eom-R} yields the algebraic approximation
\begin{equation}
    R_i \sim \frac{4c_\alpha}{M_\rho^2} \partial_j([L_j,L_i]),
\end{equation}
or, more invariantly, elimination of the heavy $\rho$-meson field generates an effective quartic pion interaction proportional to
\begin{equation}
    \Tr([L_i,L_j][L_i,L_j]).
\end{equation}
In this sense the Skyrme term may be viewed as the low-energy remnant of integrating out a sufficiently massive $\rho$-meson field.


\subsubsection{The \texorpdfstring{$\omega$}{omega}-meson variant}

A second, and historically even more striking, vector-meson extension is the Adkins--Nappi $\omega$-Skyrme model \cite{Nappi_1984}.
Here the quartic Skyrme term is omitted altogether, and stabilization is instead achieved by coupling the pion field to an isoscalar vector meson.
This is conceptually attractive because the $\omega$ field can be interpreted as the gauge field associated with the vector $\U(1)_V$ symmetry, with its coupling to the baryon current arising through the gauged Wess--Zumino term \cite{Kaymakcalan_1984}.
Unlike the $\rho$-meson, which couples to the vector pion current, the $\omega$-meson couples directly to the baryon current.
The resulting repulsion is therefore topological in character.

The fields are the non-linear $\sigma$-model field $U \in \SU(2)$ and the $\omega$-meson field $\omega_\mu \in \mathbb{R}^{1,3}$.
In physical units the Lagrangian is written as
\begin{align}
    \mathcal{L}_{\omega}[ U,\omega_\mu]  = \frac{F_{\pi}^2}{16\hbar}\,\textup{Tr}(L_{\mu} L^{\mu}) + \frac{F_{\pi}^2 m_{\pi}^2}{8\hbar^3}\,\textup{Tr}(U-\textup{Id}_2) - \frac{m_\omega^2}{2\hbar^3}\omega_\mu\omega^\mu
    - \frac{1}{4\hbar}\omega_{\mu\nu}\omega^{\mu\nu} + \beta_\omega \omega_\mu \mathcal{B}^\mu.
\end{align}
The first two terms describe the non-linear $\sigma$-model with the explicit chiral symmetry breaking pion mass potential.
A minimally broken $\U(1)_V$ Lagrangian for spin-1 mesons is given by the third and fourth terms, and the last term is the Wess--Zumino term, which describes coupling of the $\omega$-meson to three pions.
The coupling constant $\beta_\omega$ is related to the $\omega \rightarrow \pi^+ \pi^- \pi^0$ decay rate.
Here $\mathcal{B}^\mu$ is the baryon current defined in Eq.~\eqref{eq: Baryon density}.
This is the physically important distinction from the $\rho$-meson extension: the $\omega$-meson couples to the topological current rather than to the chiral current sector.

For static fields one sets $\omega_i = 0$ and writes $\omega = \omega_0$.
The static energy is then
\begin{equation}
\label{eq:omega-physical-energy}
    E_\omega[U,\omega] = \int_{\mathbb{R}^3}\d^3x \left\{ -\frac{F_\pi^2}{16\hbar}\Tr(L_iL_i) + \frac{F_\pi^2m_\pi^2}{8\hbar^3}\Tr(\Id_2-U) - \frac{1}{2\hbar}|\nabla\omega|^2 - \frac{m_\omega^2}{2\hbar^3}\omega^2 - \beta_\omega \omega \mathcal{B}^0 \right\}.
\end{equation}
Unlike the standard Skyrme energy, this functional is not bounded below if one regards $\omega$ as an unconstrained variable.
The correct static problem is instead a constrained one, because $\omega$ solves a linear elliptic equation sourced by the baryon density.
This is the technical reason the model is considerably harder to minimize numerically \cite{Sutcliffe_2009,Gudnason_2020,Leask_Harland_2024}.

\begin{figure*}[t]
    \centering
    \begin{subfigure}[b]{0.45\textwidth}
        \includegraphics[width=\textwidth]{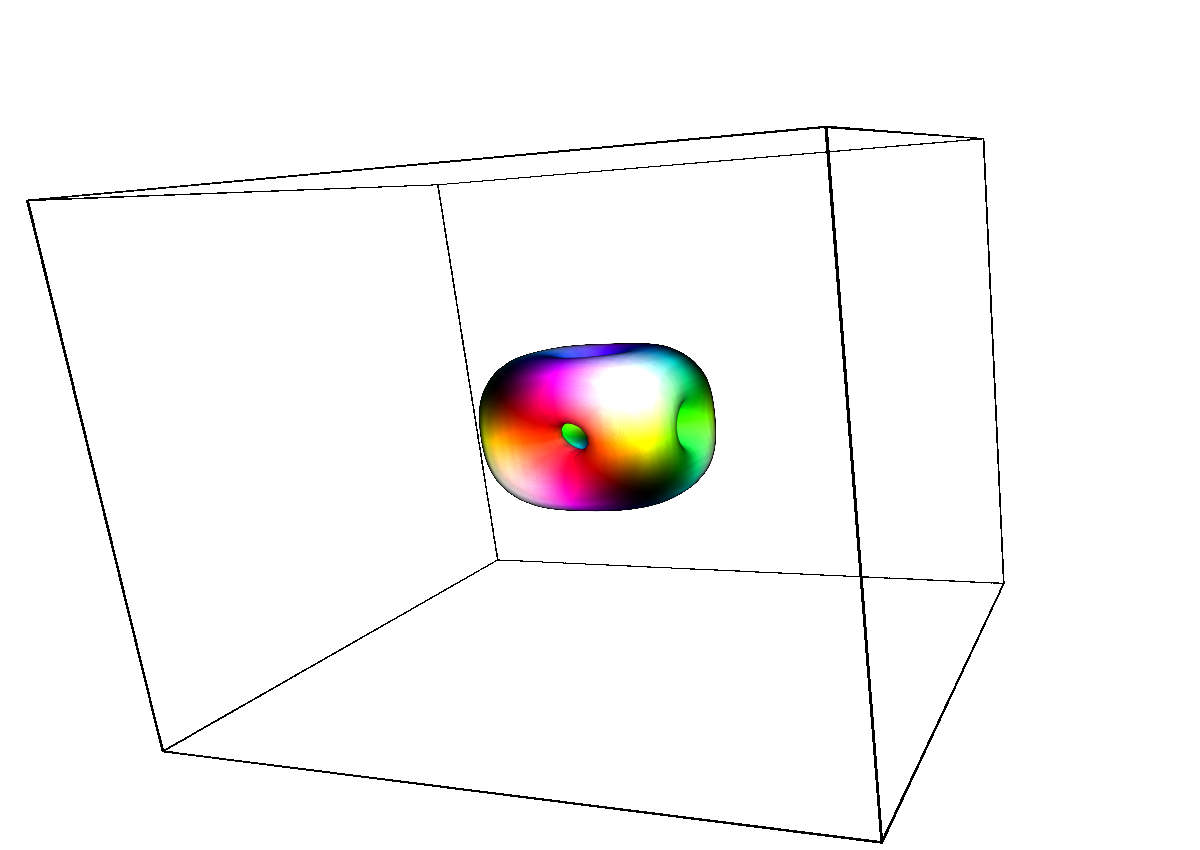}
        \caption{Baryon density}
        \label{fig: omega - runge}
    \end{subfigure}
    ~
    \begin{subfigure}[b]{0.45\textwidth}
        \includegraphics[width=\textwidth]{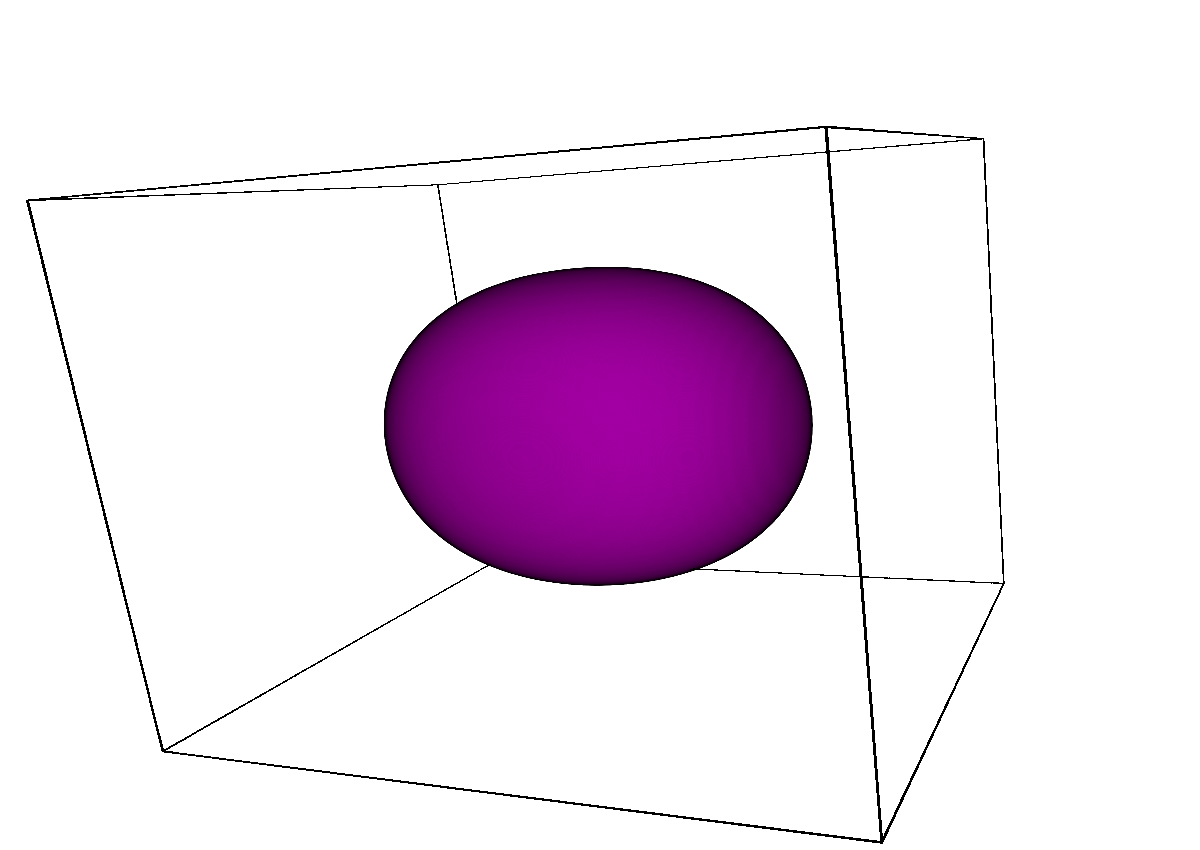}
        \caption{$\omega$-meson density, $\omega$}
        \label{fig: omega - omega}
    \end{subfigure}
    \caption{$B=4$ Skyrmion in the $\omega$-Skyrme model, showing (a) the baryon density and (b) the $\omega$-meson density.}
    \label{fig: omega}
\end{figure*}

Using the natural $\omega$-meson units $\tilde E = \frac{F_\pi^2}{m_\omega}$ and $\tilde L = \frac{\hbar}{m_\omega}$, together with the rescaling $\omega \mapsto F_\pi \omega$, one obtains the dimensionless Lagrangian
\begin{equation}
\label{eq:omega-dimensionless-lagrangian}
    \mathcal{L}_{\omega} = -\frac{m^2}{8}\Tr(\Id_2-U) + \frac{1}{16}\Tr(L_\mu L^\mu) - \frac{1}{2}\omega_\mu\omega^\mu - \frac{1}{4} \omega_{\mu\nu}\omega^{\mu\nu} + c_\omega \omega_\mu \mathcal{B}^\mu,
\end{equation}
with the dimensionless mass $m = \frac{m_\pi}{m_\omega}$ and coupling $c_\omega = \frac{m_\omega \beta_\omega}{F_\pi}$.
The associated static equations are
\begin{equation}
\label{eq:omega-static-helmholtz}
    (-\Delta + 1)\omega = - c_\omega \mathcal{B}^0,
\end{equation}
together with the pion equation
\begin{equation}
\label{eq:omega-static-eom-U}
    \partial_i L_i - \frac{m^2}{2}(U-U^\dagger) 
    + \frac{c_\omega}{\pi^2} \p_i \omega L_j L_k \epsilon^{ijk} = 0,
\end{equation}
where the last term denotes the contribution from varying the Wess--Zumino coupling $\omega \mathcal{B}^0$ with respect to $U$.
The crucial point is that the auxiliary field $\omega$ is determined by a screened Poisson, or Helmholtz, equation with source $\mathcal{B}^0$.
The cubic $B=4$ Skyrmion coupled to the $\omega$-meson is plotted in Fig.~\ref{fig: omega}.

This model also makes clear the origin of the sextic term.
In the heavy-mass limit, one neglects derivatives of $\omega$ in Eq.~\eqref{eq:omega-physical-energy}, so that the $\omega$ equation becomes algebraic:
\begin{equation}
    \omega \sim -\frac{\beta_\omega \hbar^3}{m_\omega^2}\mathcal{B}^0.
\end{equation}
Substituting back into the energy yields an effective term proportional to
\begin{equation}
    (\mathcal{B}^0)^2,
\end{equation}
that is, the sextic BPS-Skyrme term.
More precisely, one finds
\begin{equation}
    \mathcal{L}_6 = -\pi^4 \lambda^2 \eta_{\mu\nu}\mathcal{B}^\mu \mathcal{B}^\nu, \quad \lambda^2 = \frac{\beta_\omega^2 \hbar^3}{2\pi^4 m_\omega^2},
\end{equation}
so the sextic term may be interpreted as the infinite-mass limit of the $\omega$-meson field \cite{Nappi_1984,Jackson_1985}.


\subsubsection{The Coulomb-Skyrme model}

\begin{figure*}[t]
    \centering
    \begin{subfigure}[b]{0.45\textwidth}
        \includegraphics[width=\textwidth]{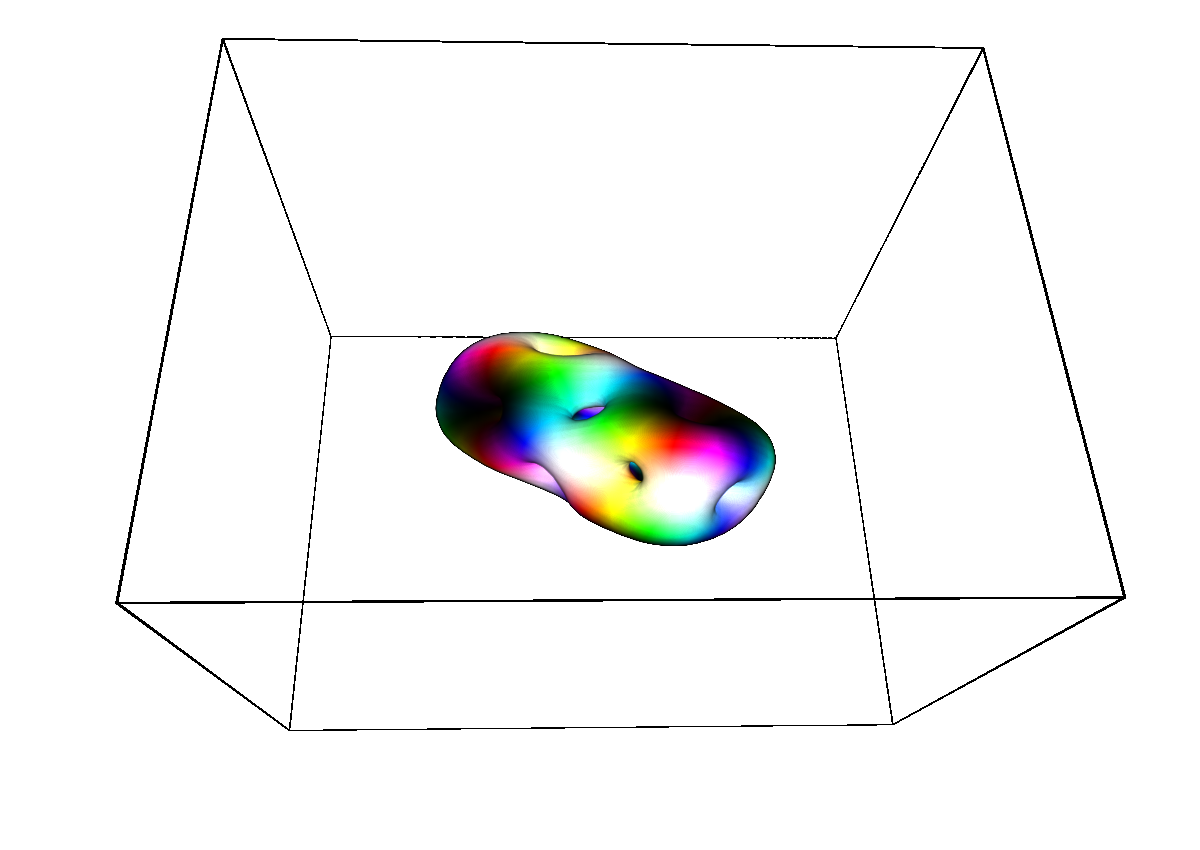}
        \caption{Baryon density}
        \label{fig: Coulomb - runge}
    \end{subfigure}
    ~
    \begin{subfigure}[b]{0.45\textwidth}
        \includegraphics[width=\textwidth]{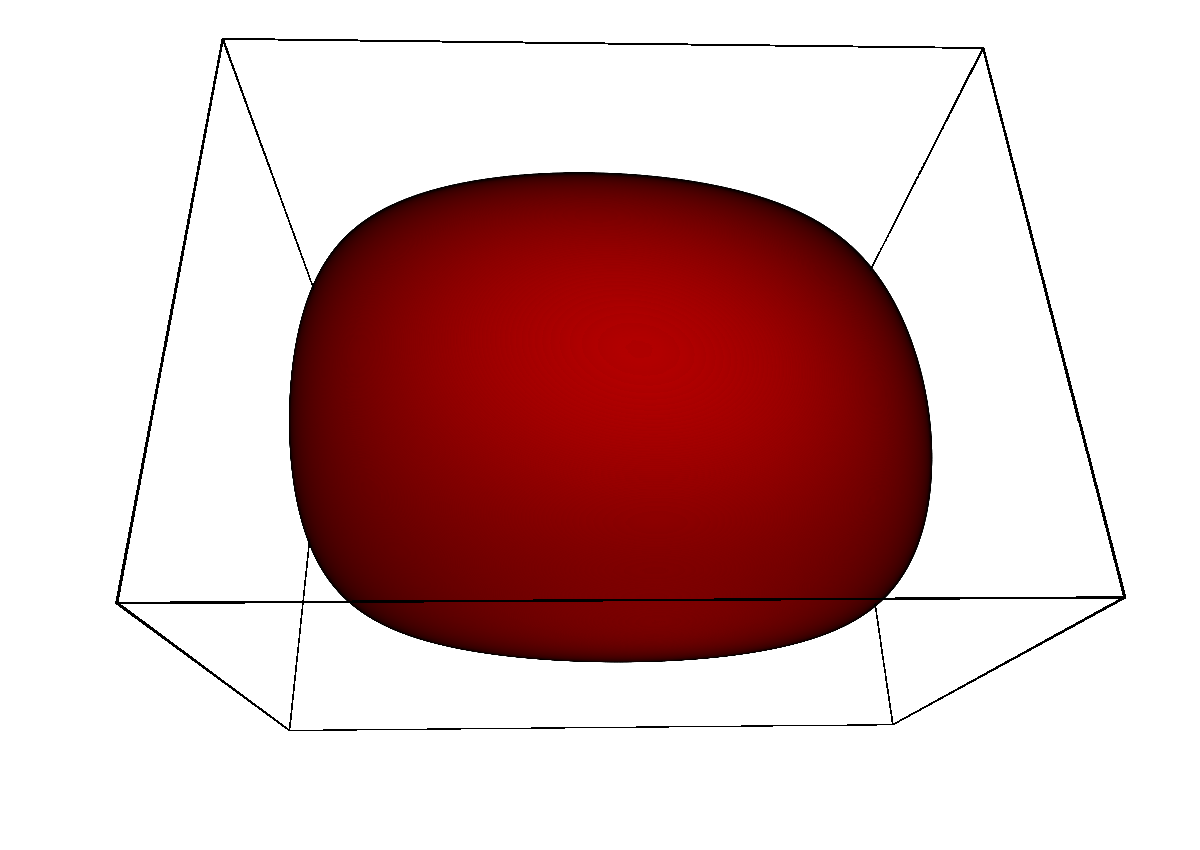}
        \caption{Coulomb potential density, $V$}
        \label{fig: Coulomb - potential}
    \end{subfigure}
    \caption{$B=8$ Skyrmion in the Coulomb-Skyrme model, showing (a) the baryon density and (b) the Coulomb potential density. Obtained using the initialization \texttt{sim.initialize(\{"mode":"smorgasbord", "baryon\_number":8, "seed":3\})}.}
    \label{fig: Coulomb}
\end{figure*}

The final built-in extension is the Coulomb-Skyrme model.
Here one augments the pion sector by an electrostatic potential and studies the backreaction of the Coulomb field on the Skyrme configuration itself \cite{Gudnason_2025}, rather than adding Coulomb energy only as a perturbative correction after solving the pion problem \cite{Ma_2019}.
For vanishing isospin states, only the electric potential survives $A_0 \in \mathbb{R}$, and the field content is $U \in \SU(2)$.
A convenient physical Lagrangian is
\begin{align}
\label{eq:coulomb-physical-lagrangian}
    \mathcal{L}_{\textup{Cou}}[U,A_0] = \, & \frac{F_\pi^2}{16\hbar}\Tr(L_\mu L^\mu) + \frac{\hbar}{32g^2}\Tr\!\left([L_\mu,L_\nu][L^\mu,L^\nu]\right) - \frac{F_\pi^2m_\pi^2}{8\hbar^3}\Tr(\Id_2-U) \nonumber \\
    \, &+ \frac{1}{2\hbar}|\nabla A_0|^2 - \frac{e}{2}A_0 \mathcal{B}^0.
\end{align}
The corresponding static energy is
\begin{align}
\label{eq:coulomb-physical-energy}
    E_{\textup{Cou}}[U,A_0] = \, & \int_{\mathbb{R}^3}\d^3x \Bigg\{-\frac{F_\pi^2}{16\hbar}\Tr(L_iL_i) - \frac{\hbar}{32g^2}\Tr\!\left([L_i,L_j][L_i,L_j]\right) + \frac{F_\pi^2m_\pi^2}{8\hbar^3}\Tr(\Id_2-U) \nonumber \\
    \, &- \frac{1}{2\hbar}|\nabla A_0|^2 + \frac{e}{2}A_0 \mathcal{B}^0 \Bigg\}.
\end{align}
Using the standard Skyrme units, $\tilde E = \frac{F_\pi}{4g}$ and $\tilde L = \frac{2\hbar}{gF_\pi}$, and introducing the rescaled variables
\begin{equation}
    V = \frac{4}{egF_\pi}A_0,
    \quad
    \kappa = \frac{e^2}{2g^2},
    \quad
    m = \frac{2m_\pi}{gF_\pi},
\end{equation}
one obtains the dimensionless static energy
\begin{align}
\label{eq:coulomb-dimensionless-energy}
    E_{\textup{Cou}}[U,V] = \, & \int_{\mathbb{R}^3}\d^3x \Bigg\{ -\frac{1}{2}\Tr(L_iL_i) - \frac{1}{16}\Tr\!\left([L_i,L_j][L_i,L_j]\right) + m^2\Tr(\Id_2-U) \nonumber \\
    \, & - \frac{\kappa}{2}(\partial_iV)^2 + \kappa V \mathcal{B}^0 \Bigg\}.
\end{align}
The static equations are
\begin{equation}
\label{eq:coulomb-poisson}
    \Delta V = \mathcal{B}^0,
\end{equation}
together with the pion equation
\begin{equation}
\label{eq:coulomb-static-eom-U}
    \partial_i \left( L_i + \frac{1}{4}[L_j,[L_i,L_j]] \right) - \frac{m^2}{2}(U-U^\dagger) 
    + \frac{\kappa}{8\pi^2} \p_i V L_j L_k \epsilon^{ijk} = 0,
\end{equation}
where the last term denotes the variation of $V\mathcal{B}^0$ with respect to $U$.
The twisted chain $B=8$ Skyrmion with the Coulomb backreaction is displayed in Fig.~\ref{fig: Coulomb}.

This theory is structurally very close to the $\omega$-Skyrme model.
In both cases the pion field is coupled, in the static problem, to a single additional scalar degree of freedom determined by an elliptic equation sourced by the baryon density.
The only essential difference is that the Coulomb field is massless and so satisfies a Poisson equation, while the $\omega$ field is massive and so satisfies a screened Helmholtz equation.
Thus the Coulomb-Skyrme and $\omega$-Skyrme models share almost the same formal structure at the level of the static variational problem, differing mainly in the presence or absence of the mass term in the auxiliary-field operator.


\subsection{Shared discretization, solver core, and observables}

The shared numerical workflow mirrors that of \texttt{cuSkyrmion}, but in a more explicitly modular form.
The solver operates on a rectangular three-dimensional lattice.
A resolved parameter set fixes the lattice dimensions, halo width, physical or dimensionless box sizes, grid spacings, number of stored field components, and solver time step before any device allocation is carried out.
These values are then packed into integer and floating-point arrays that are passed to the CUDA kernels.

The default discretization is fourth order in space, using the same type of finite-difference stencils already described in Sec.~\ref{sec: Arrested Newton flow}.
These stencils are evaluated directly in device code on flattened field buffers.

The \texttt{Simulation} class allocates the common device arrays required by the solver: fields, fictitious velocities, energy gradients, derivative work arrays, Runge--Kutta stages, scalar densities, tensor-valued observable buffers, and partial reduction arrays.
Theory-specific CUDA kernels are then injected into this common workflow (dependency injection).
The same approach is used for the physical observables.
For example, the baryon number, centre of mass, RMS radius, isospin and spin inertia tensors, mixed inertia tensor, quadrupole tensor, D-term, and virial constraint are evaluated by theory-specific kernels together with shared GPU reduction utilities.
Coordinate-dependent observables are computed in the centre-of-mass frame, exactly as in \texttt{cuSkyrmion}, in order to remove spurious translational contributions.
Their definitions need not be repeated here, since they are the same as those given in Sec.~\ref{sec: Physical properties of skyrmions}.


\subsection{Initialization procedures and interactive workflow}

The built-in initialization procedures also follow the spirit of \texttt{cuSkyrmion}.
For pion-only theories, the package supports hedgehog and rational-map ans\"atze together with product constructions for multi-Skyrmion initial states.
A stochastic initialization mode in the style of the sm\"org\aa sbord method \cite{Gudnason_2022} is also included, in which randomly positioned and randomly oriented unit Skyrmions are combined into a composite initial configuration before relaxation.
The underlying ideas are the same as those described in Sec.~\ref{sec: Constructing multiskyrmion configurations}, but are exposed here through a theory registry and a common \texttt{Simulation} interface rather than through a single hard-wired code path.

\begin{figure*}[t]
    \centering
    \includegraphics[width=0.7\textwidth]{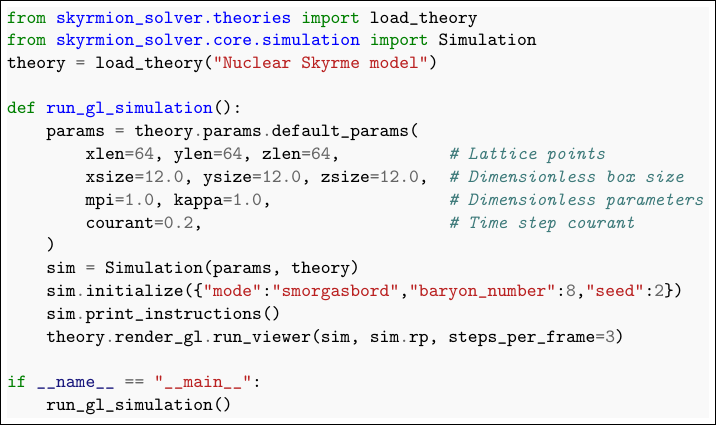}
    \caption{Typical usage of \texttt{skyrmion\_solver}. A theory is loaded from the registry, parameters are resolved, a \texttt{Simulation} object is constructed, an initial configuration is generated, and the GPU-resident viewer is launched. The same workflow applies across the built-in Skyrme-type models.}
    \label{fig: Usage}
\end{figure*}

A typical workflow is shown schematically in Fig.~\ref{fig: Usage}.
A theory is loaded from the registry, the corresponding parameters are resolved, a \texttt{Simulation} object is constructed, an initial configuration is generated, and the GPU-resident viewer is launched.
From that point onward, the user can run arrested Newton flow or explicit RK4 evolution, inspect the result in real time, and query observables during the run.
The user-facing interface is intentionally stable across theories: the main changes from one model to another are the available parameters, initialization options, and observables exposed by the theory module.

Overall, \texttt{skyrmion\_solver} should therefore be viewed not simply as a \texttt{Python} rewrite of \texttt{cuSkyrmion}, but as a modular CUDA framework for three-dimensional baryonic solitons.
It preserves the low-level GPU execution model of the original code, including explicit kernels, low-level launches, SIMT execution, and CUDA--OpenGL interoperability, while making the surrounding software architecture substantially more reusable and extensible.
A typical workflow for \texttt{skyrmion\_solver} is shown in Fig.~\ref{fig: Usage} for the standard massive nuclear Skyrme model.

\section{Outlook}
\label{sec: Outlook}

The \texttt{cuSkyrmion} code is designed to separate numerical evolution and visualization. 
This modular structure facilitates extension to additional Skyrme-like models, alternative observables, and larger lattice simulations.
The computational module is not itself modular on the level of specifying the theory and using the same computational core.
We have developed also a \texttt{Python}-fork of the programme, \texttt{skyrmion\_solver}, that utilizes the rendering kernels from \texttt{cuSkyrmion} \emph{mutatis mutandis}.
This demonstrates that the rendering/visualization module can readily be used by other computational modules as written.
We have also demonstrated that \texttt{skyrmion\_solver} is readily generalizable to a vast landscape of Skyrmion-like models with extra field content and couplings.

Extensions to a multi-GPU version or a framework with a distributed
execution of the computation module coupled with a visualization
module on one node would both be interesting, which we leave for
future work.


\subsection*{Acknowledgements}
S.~B.~G.~thanks the Outstanding Talent Program of Henan University for partial support.
P.~L. acknowledges funding from the Olle Engkvists Stiftelse through the grant 226-0103.

\appendix

\bibliography{bib.bib}

\end{document}